\documentclass[sigconf]{acmart}
\renewcommand\footnotetextcopyrightpermission[1]{} 
\AtBeginDocument{%
  \providecommand\BibTeX{{%
    \normalfont B\kern-0.5em{\scshape i\kern-0.25em b}\kern-0.8em\TeX}}}

\usepackage{algorithmic}
\usepackage{graphicx}
\usepackage{textcomp}
\usepackage{xcolor}

\usepackage{subfiles} 
\usepackage{multirow}
\usepackage{listings}
\usepackage{subcaption}
\usepackage{color,xcolor,colortbl}
\usepackage{xspace}
\usepackage{enumitem}
\usepackage{graphicx}
\usepackage{xr}
\usepackage{url}

\usepackage[most]{tcolorbox} 
\usepackage{booktabs}
\usepackage{diagbox}
\usepackage{amsmath}
\usepackage{array}
\usepackage{colortbl}
\usepackage{color}
\usepackage{balance}
\usepackage{arydshln}
\usepackage{tabularx}

\newcommand\clearrow{\global\let\rowmac\relax}
\clearrow
 \usepackage{adjustbox}
\usepackage[flushleft]{threeparttable}
\usepackage{amsmath, amsfonts}
\usepackage{algorithmic}
\usepackage{graphicx}
\usepackage{textcomp}
\usepackage{xcolor}
\usepackage{xspace}
\usepackage{booktabs}
\usepackage{multirow}
\usepackage{subcaption}
\usepackage{enumitem}
\usepackage{graphicx}  %
\usepackage{microtype}
\usepackage{balance}
\usepackage{soul}

\usepackage{hyperref}
\usepackage{hyperxmp}
\usepackage{cleveref}

\graphicspath{ {./figures/} }
\AtBeginDocument{%
  \providecommand\BibTeX{{%
    \normalfont B\kern-0.5em{\scshape i\kern-0.25em b}\kern-0.8em\TeX}}}

\def\BibTeX{{\rm B\kern-.05em{\sc i\kern-.025em b}\kern-.08em
    T\kern-.1667em\lower.7ex\hbox{E}\kern-.125emX}}

\definecolor{lightgreen}{rgb}{0.56, 0.93, 0.56}
\definecolor{magicmint}{rgb}{0.67, 0.94, 0.82}
\definecolor{mintgreen}{rgb}{0.6, 1.0, 0.6}
\definecolor{mediumspringgreen}{rgb}{0.0, 0.98, 0.6}
\definecolor{lightgray}{HTML}{EFEFEF}

\begin{document}

\title{Comparison of Static Application Security Testing Tools and Large Language Models for Repo-level Vulnerability Detection}

\author{Xin Zhou}
\affiliation{%
  \institution{Singapore Management University}
  \country{Singapore}}
\email{xinzhou.2020@phdcs.smu.edu.sg}

\author{Duc-Manh Tran}
\affiliation{%
  \institution{Hanoi University of Science and Technology}
  \country{Vietnam}}
\email{manh.td120901@gmail.com}

\author{Thanh Le-Cong}
\affiliation{%
  \institution{The University of Melbourne}
  \country{Australia}}
\email{congthanh.le@student.unimelb.edu.au}

\author{Ting Zhang}
\affiliation{%
  \institution{Singapore Management University}
  \country{Singapore}}
\email{tingzhang.2019@phdcs.smu.edu.sg}

\author{Ivana Clairine IRSAN}
\affiliation{%
  \institution{Singapore Management University}
  \country{Singapore}}
\email{ivanairsan@smu.edu.sg}

\author{Joshua SUMARLIN}
\affiliation{%
  \institution{Singapore Management University}
  \country{Singapore}}
\email{jsumarlin.2022@scis.smu.edu.sg}

\author{Bach Le}
\affiliation{%
  \institution{University of Melbourne}
  \country{Australia}}
\email{bach.le@unimelb.edu.au}

\author{David Lo}
\affiliation{%
  \institution{Singapore Management University}
  \country{Singapore}}
\email{davidlo@smu.edu.sg}

\begin{abstract}

Software vulnerabilities pose significant security challenges and potential risks to society, necessitating extensive efforts in automated vulnerability detection.
There are two popular lines of work to address automated vulnerability detection. On one hand, Static Application Security Testing (SAST) is usually utilized to scan source code for security vulnerabilities, especially in industries. On the other hand, deep learning (DL)-based methods, especially since the introduction of large language models (LLMs), have demonstrated their potential in software vulnerability detection. However, there is no comparative study between SAST tools and LLMs, aiming to determine their effectiveness in vulnerability detection, understand the pros and cons of both SAST and LLMs, and explore the potential combination of these two families of approaches.

In this paper, we compared 15 diverse SAST tools with 12 popular or state-of-the-art open-source LLMs in detecting software vulnerabilities from repositories of three popular programming languages: Java, C, and Python. The experimental results showed that SAST tools obtain low vulnerability detection rates with relatively low false positives, while LLMs can detect up 90\% to 100\% of vulnerabilities but suffer from high false positives.
By further ensembling the SAST tools and LLMs, the drawbacks of both SAST tools and LLMs can be mitigated to some extent.
Our analysis sheds light on both the current progress and future directions for software vulnerability detection.

\end{abstract}

\maketitle

\vspace{-0.2cm}
\section{Introduction}

The early detection and mitigation of software vulnerabilities is crucial, as unaddressed vulnerabilities can potentially lead to consequences like massive economic losses and compromised critical infrastructure.
With the number of software vulnerabilities significantly increasing from 5,697 in 2013 to 29,065 in 2023~\cite{CVE_NUM}, the impact of software vulnerabilities has also amplified.
Accurately and promptly identifying vulnerabilities is crucial for mitigating potential risks and has garnered significant attention from both industry and academia.

Static Application Security Testing (SAST) tools are commonly used in industries to scan source code and detect vulnerabilities. These tools analyze code without executing it, using techniques such as syntax, data flow, and control flow analysis~\cite{croft2021empirical}. SAST is popular due to its low cost, fast operation, and ability to find bugs without running the program~\cite{Java_CVE}.
Meanwhile, in the research community, large language model (LLM)-based methods have demonstrated their effectiveness for function-level software vulnerability detection~\cite{DBLP:conf/msr/FuT22/linevul,liu2024pre}, which predict whether a function contains the vulnerability or not.
LLMs are deep learning (DL) models based on the Transformer architecture~\cite{vaswani2017attention}, comprising a vast number of parameters and pre-trained on massive amounts of source code, text, and other data modalities~\cite{gpt3,codellama}. 
Recently, LLMs have seen a significant increase in the size of their pre-training data and model parameters~\cite{llama3,starcoder2}. The extensive knowledge acquired through the large-scale pre-training offers the potential to substantially improve the effectiveness of vulnerability detection.

Despite the considerable research interest in either SAST tools or LLMs for vulnerability detection, a comprehensive comparative study between these two popular approaches has been lacking. This gap can be attributed to the following three major \textbf{challenges}:

\textbf{Lack of a repo-level vulnerability detection formulation:}
SAST tools are usually used to detect vulnerabilities across entire code repositories. However, current DL-based methods for vulnerability detection, including those utilizing LLMs, primarily focus on identifying vulnerabilities within individual functions~\cite{devign,ReVeal,codebert,liu2024pre, DBLP:conf/icse/YangWLW23}. To date, we are unaware of any DL-based methods that are designed to detect vulnerabilities at the level of entire repositories. This gap in methodology hinders a fair and direct comparison between SAST tools and LLMs. 
Therefore, there is a need for a new task formulation for repo-level vulnerability detection.

\textbf{Lack of datasets supporting both SAST and LLMs:}
Existing datasets for DL-based or LLM-based vulnerability detection, such as BigVul~\cite{BigVul} and CVEfixes~\cite{CVEfixes}, contain labeled data at the function level without the source code of complete repositories, which is not suitable for conducting experiments at the repository level. 
Thus, there is a need for datasets that contain the complete source code of vulnerable repositories, which are suitable for evaluating both SAST and LLM-based approaches.

\textbf{Lack of a comprehensive LLM evaluation framework:} 
Recently, there has been a proliferation of diverse LLMs~\cite{Huggingface_OpenLLM}, and their implementations could vary in terms of model architectures (i.e., encoder-only, decoder-only,
encoder-decoder). Moreover, adapting these LLMs to specific downstream tasks, such as repo-level vulnerability detection, often could employ various adaptation techniques. These techniques include zero-shot prompting~\cite{gpt3}, few-shot prompting~\cite{gpt3}, chain-of-thought prompting~\cite{cot}, and fine-tuning~\cite{lora,codebert}, etc.
Given the complexity and diversity of LLMs, there is a need for a unified LLM evaluation framework that can support the diverse LLMs and various adaptation techniques.

\begin{figure}[t]
    \centering
    \includegraphics[width=0.5\textwidth]{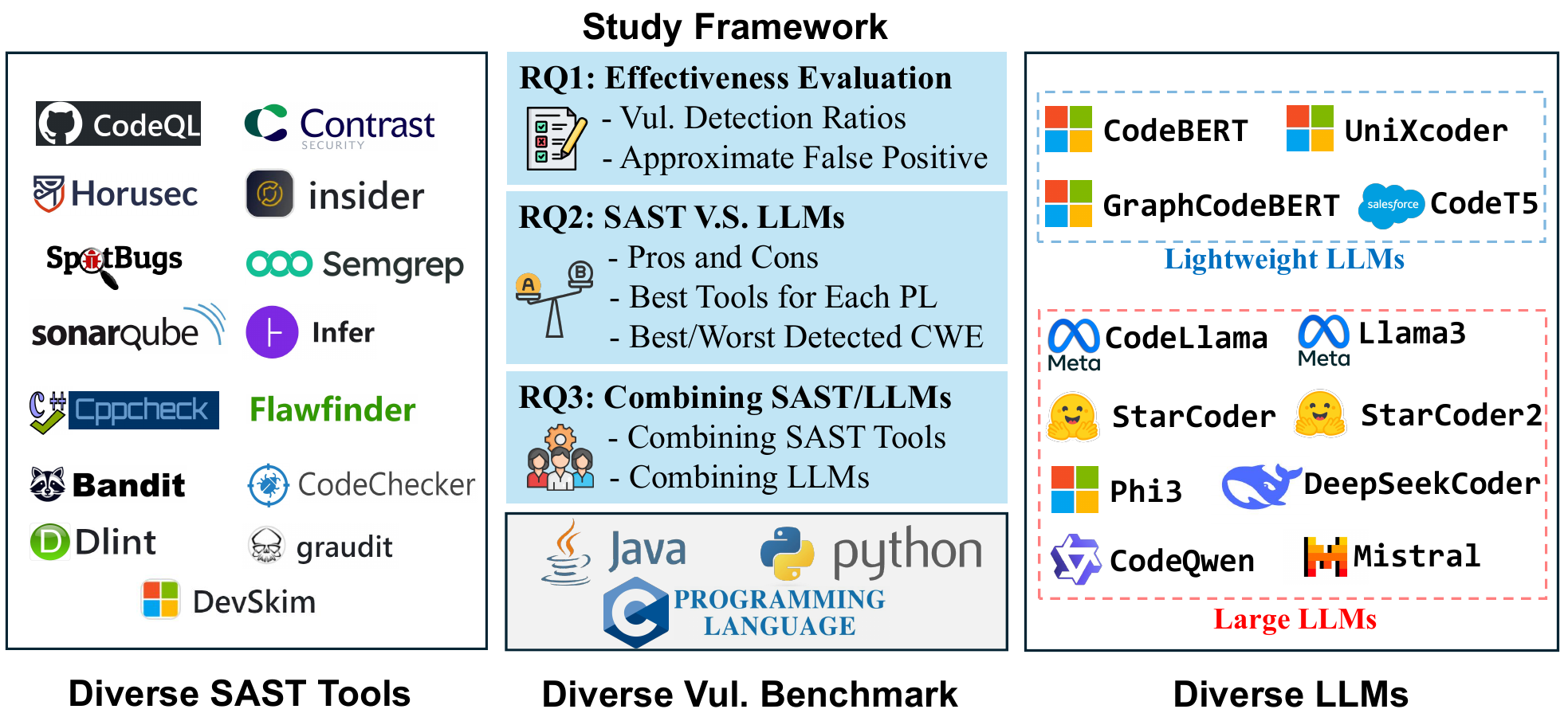}
    \vspace{-0.7cm}
    \caption{Overview of Our Study}
  \label{fig:framework}
  \vspace{-0.65cm}
\end{figure}

To tackle these challenges, in this paper, we conduct a comprehensive study as shown in Figure~\ref{fig:framework}, aimed at comparing SAST tools with LLMs in detecting software vulnerabilities from repositories.
Firstly, we formulate a novel task that enables LLMs to conduct repo-level vulnerability detection. The task involves identifying the vulnerability from a given software repository.
Secondly, we have curated three datasets for repo-level vulnerability detection, each tailored to a widely used programming language: Java, C, and Python. 
This diverse set of evaluation datasets enable us to gain insights into the proficiency of SAST tools and LLMs in detecting vulnerabilities across these different popular programming languages.
Thirdly, we implement a comprehensive evaluation framework encompassing 12 distinct LLMs with diverse architectures (encoder-only, decoder-only, encoder-decoder) and parameter sizes ranging from 120 million to 8 billion parameters.

Our experiments yield the following key \textbf{findings}:
\noindent
\begin{enumerate}[leftmargin=*,noitemsep,nolistsep]
\item SAST tools obtained low vulnerability detection ratios (up to 44.4\%) and many failed to detect any vulnerabilities. 
\item LLMs achieved high vulnerability detection rates (up to 90\% to 100\%) but they were accompanied by high false positive rates.
\item By combining multiple SAST tools, we observed a substantial increase (i.e., 25.2\%--100.0\%) in vulnerability detection rates.
Combining diverse LLMs led to a substantial reduction (i.e., 40.9\%--74.6\%) in false positive rates. 
\item The best approach differs across programming languages if considering both detection rates and false positive rates. For Java, the combined LLMs provide the most effective solution, while for C and Python, the combined SAST tools are the best.
\end{enumerate}

\vspace{0.1cm}
\noindent\textbf{Contributions.} 
In summary, our contributions are as follows:
\begin{itemize} [leftmargin=*]
\item [$\bullet$] To the best of our knowledge, we are the first to comprehensively compare the SAST tools with LLMs in three popular programming languages, Java, C, and Python.
\item [$\bullet$] We are the first to introduce the task of repo-level vulnerability detection and evaluate twelve diverse LLMs on this task. 
\item [$\bullet$] We have constructed a new repo-level Python vulnerability dataset and curated Java and C vulnerability datasets, facilitating evaluation across multiple popular programming languages.
\item [$\bullet$] As far as we know, we are the first to evaluate the SAST tools in detecting real-world vulnerabilities in Python repositories.
\item [$\bullet$] We conduct an extensive evaluation of 15 SAST tools and 12 LLMs, including four different LLM adaptation techniques. We identify the best-performing approaches and discuss implications for future studies on vulnerability detection.
\end{itemize}

\vspace{-0.2cm}
\section{Study Design}
In this section, we present our proposed repo-level vulnerability detection task. We then introduce the data collection and detail the study objects (i.e., SAST tools and LLMs).

\vspace{-0.2cm}
\subsection{Repo-level Vulnerability Detection}
We first introduce our proposed repo-level vulnerability detection task and then compare it with the function-level vulnerability detection task used in existing studies~\cite{devign,codebert,ReVeal}.

\vspace{-0.1cm}
\subsubsection{Repo-level Vulnerability Detection Task Formation}
This task aims to detect the vulnerable functions in an entire repository. It aims to learn a detector $M$ that can map the source code of an entire repository to vulnerability labels at the function level:
$$M:\mathcal{X}\mapsto\mathcal{Y}, \;\; \mathcal{X}=\{x_0,x_1,...,x_n\} \;\text{and}\; \mathcal{Y}=\{0,1,...,0\}$$
where $\mathcal{X}$ denotes the entire input repository and its component $x_i$ refers to the $i$-th function in the input repository. $n$ refers to the number of functions in the input repository. $\mathcal{Y}$ denotes the labels of all functions in the repository, where 1 indicates a vulnerable function and 0 indicates a clean (non-vulnerable) function.

\vspace{-0.1cm}
\subsubsection{Comparison with Function-level Vulnerability Detection}
The function-level vulnerability detection aims to predict whether a target function contains a vulnerability or not~\cite{devign,ReVeal}. To understand the difference between function-level and repo-level vulnerability detection, we need to briefly explain what the ``target functions'' are in the function-level vulnerability detection dataset/task.

\begin{figure}[t]
    \centering
    \includegraphics[width=0.5\textwidth]{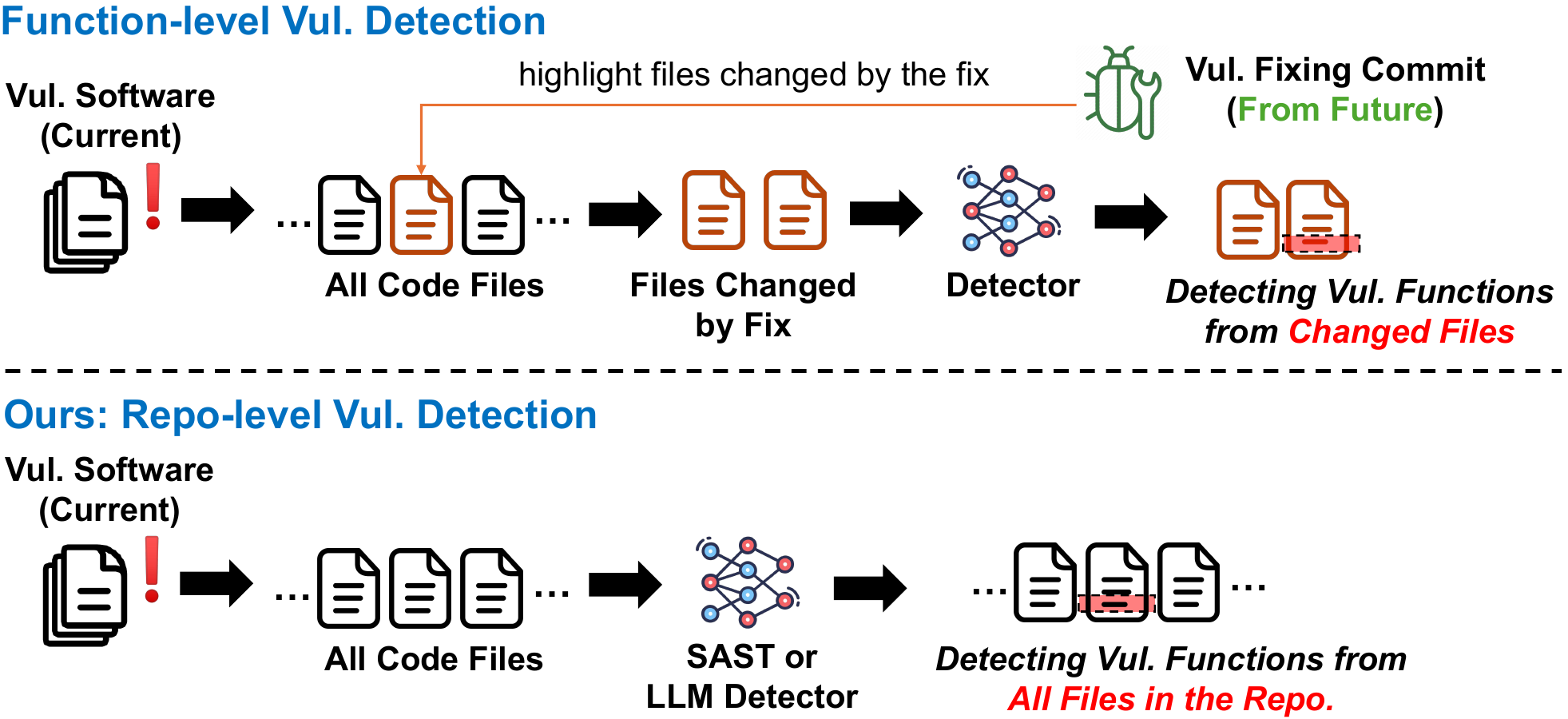}
    \vspace{-0.7cm}
    \caption{Repo-level and Function-level Vul. Detection}
  \label{fig:repo-fun-compare}
  \vspace{-0.5cm}
\end{figure}

Many function-level vulnerability detection datasets, such as BigVul~\cite{BigVul}, ReVeal~\cite{ReVeal}, CrossVul~\cite{crossvul}, CVEfixes~\cite{CVEfixes}, and DiverseVul~\cite{chen2023diversevul}, employ a heuristic to automatically label functions, as depicted in Figure~\ref{fig:repo-fun-compare}. Firstly, they collect vulnerability-fixing commits from databases like NVD. 
They then label the pre-commit versions of functions changed by the vulnerability-fixing commits as vulnerable functions.
Additionally, they label all unchanged functions in the changed code files in vulnerability-fixing commits as non-vulnerable. Thus, the ``target functions'' are functions in the changed code files of vulnerability-fixing commits.
As shown in the bottom part of Figure~\ref{fig:repo-fun-compare}, in contrast, our proposed repo-level vulnerability detection aims to detect vulnerable functions given all functions in the repository. Thus, the repo-level task has a much broader detection scope.
For instance, the ratios of vulnerable functions in the function-level vulnerability detection datasets BigVul~\cite{BigVul}, ReVeal~\cite{ReVeal}, CrossVul~\cite{crossvul}, CVEfixes~\cite{CVEfixes}, and DiverseVul~\cite{chen2023diversevul} range from 4.5\% to 9.2\%. In contrast, the vulnerable function ratios in our studied repo-level vulnerability detection datasets are substantially lower, ranging from 0.008\% to 0.18\% as shown later in Table~\ref{tab:dataset}. 

We propose the repo-level vulnerability detection task for two reasons: (1) The repo-level task is more aligned with the tasks performed by SAST tools, which usually check the entire repository to detect vulnerabilities; (2) Limiting the detection range to only the changed files in vulnerability-fixing commits is not a practical approach for vulnerability detection. 
This is because when vulnerabilities are yet to be discovered and addressed, the corresponding vulnerability-fixing commits do not exist. Therefore, it is better not rely on such commits to narrow down the detection scope from the entire repository into files changed in vulnerability-fixing commits.

\vspace{-0.1cm}
\subsubsection{Applying LLMs for Repo-level Vulnerability Detection}
For the repo-level vulnerability detection task, the goal is to detect vulnerable functions across all functions within an entire repository. However, repositories can be extremely lengthy, while some of the studied LLMs, such as CodeBERT and GraphCodeBERT, have input length limitations of 512 tokens. This 512-token limit aligns well with function-level data but not with classes or entire repositories.
Although some larger studied LLMs allow wider input ranges (e.g., up to 8,192 tokens for StarCoder~\cite{starcoder}), we aim to conduct experiments within the same setting for all studied LLMs. To enable all LLMs, including LLMs like CodeBERT, to perform the repo-level task, we employ the following strategy unified across all LLMs:
(1) We split a repository into individual functions.
(2) We predict the vulnerability label for each function in the repository separately.
(3) We aggregate all predictions for the functions within the repository to obtain the predictions for the entire repository.
(4) We analyze whether vulnerabilities are detected within the repository based on the aggregated predictions (See Sections \ref{DetectionScenarios} and \ref{EvaluationMetrics} for details).

\vspace{-0.2cm}
\subsection{Vulnerability Datasets}

In this paper, we investigate three datasets for three popular programming languages, i.e., Java, C, and Python. Specifically, the Java and C datasets are sourced from Li et al.~\cite{Java_CVE} and Lipp et al.~\cite{C_CVE}. 
The Python dataset is newly constructed.
The statistics of three datasets are shown in Table~\ref{tab:dataset}.
Please note that for each CVE entry, our datasets provided 1) CVE ID, 2) CWE type, 3) the complete source code of the corresponding vulnerable software, and 4) locations of vulnerabilities at the function level.

\vspace{0.05cm}
\noindent\textbf{Java and C Vulnerability Datasets.}
For the Java vulnerability dataset selection, we considered the Java dataset shared by Li et al.~\cite{Java_CVE}, as it contains real-world vulnerabilities, in contrast to synthetic datasets such as the Juliet Test Suite Java~\cite{Juliet}. This dataset contains CVE IDs, project names, and the locations of vulnerabilities labeled at the function level. However, the authors of the dataset did not share the actual repositories of the vulnerable Java software. To obtain the complete source code of vulnerable repositories, we searched for the vulnerable repositories with the provided version IDs and project names on GitHub.
We only selected the repositories that we could successfully download through GitHub APIs. 
In total, we obtained 87 CVE entries and the corresponding vulnerable repositories.
Next, we utilized Tree-sitter~\cite{Tree-sitter} to parse and slice each vulnerable repository into individual function-level code snippets.
For the C vulnerability dataset selection, we considered the dataset shared by Lipp et al.~\cite{C_CVE} which contains real-world vulnerabilities. 
Similarly, this dataset contains CVE IDs, and project names but not the actual vulnerable C repositories. To obtain the complete source code of vulnerable repositories, we followed the instructions of the authors and successfully obtained the vulnerable repositories of 85 CVE entries. 
Next, we utilized Tree-sitter~\cite{Tree-sitter} to parse and slice each vulnerable repository into individual function-level code snippets.

\begin{table}[t]
\caption{Statistics of the Used Datasets.}
\vspace{-0.45cm}
\resizebox{0.4\textwidth}{!}{
\begin{tabular}{c|c|c|c|c}
\hline
\multicolumn{1}{l|}{\textbf{Dataset}} & \textbf{\#Project} & \textbf{\#CVE} & \textbf{\#Functions} & \textbf{Vul. Ratio} \\ \hline
Java                                  & 50                 & 87             & 3,986,848            & 0.008\%            \\ \hline
C                                     & 6                  & 85             & 1,181,556            & 0.009\%            \\ \hline
Python                                & 78                 & 98             & 312,045              & 0.18\%             \\ \hline
\end{tabular}
}
\label{tab:dataset}
\vspace{-0.6cm}
\end{table}

\vspace{0.05cm}
\noindent\textbf{Python Vulnerability Dataset.}
Python is one of the most popular programming languages~\cite{python_popularity}, with a broad range of application scenarios. 
However, there has been a lack of research effort in two aspects. 
Firstly, although previous studies\cite{C_CVE,Java_CVE} have evaluated SAST tools for C and Java respectively, their findings may not be directly applicable to Python due to the differences in language semantics and characteristics.
Secondly, other studies on applying LLMs to detect vulnerabilities have primarily focused on C/C++~\cite{devign,BigVul,CVEfixes}, 
leaving a research gap for Python.
To address this gap, we have constructed a new dataset of real-world Python vulnerabilities.

We first downloaded all available vulnerability data in JSON format from the National Vulnerability Database (NVD) from 2002 to 2023. We then parsed this data and extracted GitHub Python repositories by collecting the reference URLs of these vulnerabilities. We then identified the fixing commits for each vulnerability by checking the reference links provided in the vulnerability reports. 
Subsequently, we selected the repositories from which we can retrieve complete source code via GitHub.
We omitted projects that encountered errors during scanning with SAST tools.
Through these steps, we curated a dataset comprising a total of 98 CVE entries related to Python projects hosted on GitHub.

To identify the vulnerable functions from vulnerability-fixing commits, we adopted the approach proposed by previous methods~\cite{BigVul,CVEfixes,devign}. First, we collected the software versions prior to each vulnerability-fixing commit, which are regarded as vulnerable versions. Then, we labeled the functions containing lines of code that were modified in the fixing commit as vulnerable. All other functions present in the vulnerable version were considered non-vulnerable.
Although these other functions could potentially harbor vulnerabilities that have not been discovered by the community, in this work, we aim to assess whether the SAST tools/LLMs could accurately identify known vulnerabilities. Therefore, we assume that it is reasonable to label all functions, except the known vulnerable ones, as non-vulnerable.

\vspace{-0.3cm}
\subsection{SAST Tools Selection}

\noindent\textbf{Java and C SAST Tools.}
For Java SAST tools, we followed Li et al.~\cite{Java_CVE} to select 7 free or open-source SAST tools that support Java code: CodeQL~\cite{CodeQL}, Contrast Codesec Scan (Contrast)~\cite{Contrast}, Horusec~\cite{Horusec}, Insider~\cite{Insider}, SpotBugs~\cite{SpotBugs}, Semgrep~\cite{Semgrep}, and SonarQube community edition (SonarQube)~\cite{SonarQube}. These tools encompass a diverse range of SAST techniques and are popular among developers, as indicated by the number of GitHub stars~\cite{Java_CVE}.
For C SAST tools, we followed Lipp et al.\cite{C_CVE} to select five free or open-source SAST tools that support C code, i.e., Flawfinder~\cite{Flawfinder}, Cppcheck~\cite{Cppcheck}, Infer~\cite{Infer}, CodeChecker~\cite{CodeChecker}, and CodeQL~\cite{CodeQL}.
These tools implemented state-of-the-art analysis techniques~\cite{C_CVE} and were used in previous studies focusing on evaluating SAST tools~\cite{C_CVE,stefanovic2020static}.

\vspace{0.05cm}
\noindent\textbf{Python SAST Tools.}
We aim to select a representative set of SAST tools that support Python code.
For the search range, we utilized a comprehensive set of 576 candidate SAST tools identified by Li et al.~\cite{Java_CVE}. They compiled their list by searching tool lists from recent literature (e.g., ~\cite{aloraini2019empirical,goseva2015capability,liu2023comprehensive}) and obtaining recommendations for SAST tools from several prominent websites (e.g., GitHub, NIST, and Wikipedia).
We designed the following criteria to select SAST tools for Python:
1) Python supported: we included only SAST tools that support Python programs;
2) Free of charge: due to the substantial costs of commercial tools, we followed Li et al.~\cite{Java_CVE} to select the free Python SAST tools;
3) Security-related: we selected tools that identify generalized security vulnerabilities rather than those aimed at detecting specific vulnerabilities or code quality issues;
4) Well-documented with detecting rules: 
 we selected SAST tools with clear documentation.
Based on these criteria, we finally selected six SAST tools: Bandit~\cite{bandit}, Dlint~\cite{Dlint}, DevSkim~\cite{DevSkim}, CodeQL~\cite{CodeQL}, Graudit~\cite{graudit}, and Semgrep~\cite{Semgrep}.

\vspace{-0.2cm}
\subsection{Studied LLMs}
We focus on free and open-source LLMs instead of commercial LLMs such as GPT-3.5 and GPT-4 for two primary reasons. Firstly, the cost of utilizing commercial LLMs to scan entire repositories for vulnerability detection is prohibitively high. As illustrated in Table~\ref{tab:dataset}, the datasets contain 312,045 to 3,986,848 functions, making the use of commercial LLMs financially challenging for such large-scale tasks. 
Secondly, commercial LLMs undergo continuous updates, making it difficult to replicate and validate the results. In contrast, open-source LLMs offer greater transparency and accessibility, facilitating reproducible evaluations.

Table~\ref{tab:studied_llm} summarizes the characteristics of the studied LLMs.
Please note that in accordance with a recent comprehensive survey on large language models for Software Engineering~\cite{hou2024large}, we categorize CodeBERT, GraphCodeBERT, CodeT5, and UniXcoder as LLMs, though their model sizes may be considered relatively small within the LLM spectrum.
Specifically, we select two groups of open-source LLMs: 1) general LLMs and 2) code-related LLMs.
General LLMs are pre-trained on diverse data, including natural language and code, and can be utilized for various tasks. In contrast, code-related LLMs are specifically (further) pre-trained on code-related data. Given the empirical nature of this study, we aim to assess the effectiveness of both LLM categories in the repo-level vulnerability detection task.
For code-related LLMs, we select the five recently released and leading models (in 2023 or 2024): DeepSeek-Coder (DSCoder)~\cite{deepseekcoder}, CodeLlama~\cite{codellama}, StarCoder~\cite{starcoder}, StarCoder2~\cite{starcoder2}, and CodeQwen~\cite{codeqwen}. Additionally, we select four popular and lightweight code-related LLMs with parameters less than 1 billion: CodeBERT~\cite{codebert}, GraphCodeBERT (G-BERT)~\cite{graphcodebert}, CodeT5~\cite{codet5}, and UniXcoder~\cite{unixcoder}.
Regarding general LLMs, we select three recently released models: Mistral~\cite{mistral}, Phi3~\cite{phi3}, and Llama3~\cite{llama3}.
Since LLM families have different model sizes, we select models with no more than \textbf{8B parameters}. 
This constraint on the model parameters is imposed by our computing resources (NVIDIA RTX A5000).

\begin{table}[t]
\centering
\caption{Overview of Studied LLMs.}
\vspace{-0.4cm}
\resizebox{0.49\textwidth}{!}{
\begin{tabular}{@{}llrrrcc@{}}
\toprule
 & \textbf{Model} & \textbf{\#Param.} & \textbf{Type} & \textbf{\begin{tabular}[c]{@{}r@{}}Release \\ Date\end{tabular}} & \textbf{\begin{tabular}[c]{@{}c@{}}Code Gene.\\ (HumanEval) \\ Pass@1\end{tabular}} & \textbf{\begin{tabular}[c]{@{}c@{}}Adaptation\\ Techniques\end{tabular}} \\ \midrule
\multicolumn{1}{l|}{\multirow{4}{*}{\textbf{\begin{tabular}[c]{@{}l@{}}Light-\\ weight\\ LLMs \\ ($<$1B)\end{tabular}}}} & CodeBERT & 0.13B & Enc & Jul. 2020 & - & \multirow{4}{*}{\begin{tabular}[c]{@{}c@{}}Full \\ Fine-tuning\end{tabular}} \\
\multicolumn{1}{l|}{} & G-BERT & 0.13B & Enc & Feb. 2021 & - &  \\
\multicolumn{1}{l|}{} & CodeT5 & 0.22B & Enc-Dec & Sep. 2021 & - &  \\
\multicolumn{1}{l|}{} & UniXcoder & 0.13B & Enc-Dec & Mar. 2022 & - &  \\ \midrule
\multicolumn{1}{l|}{\multirow{8}{*}{\textbf{\begin{tabular}[c]{@{}l@{}}Large\\ LLMs \\ ($\geq$1B)\end{tabular}}}} & StarCoder & 7B & Dec & May 2023 & 20.4\% & \multirow{8}{*}{\begin{tabular}[c]{@{}c@{}}PEFT,\\ Zero-shot Prompt,\\ Few-shot Prompt,\\ CoT Prompt\end{tabular}} \\
\multicolumn{1}{l|}{} & CodeLlama & 7B & Dec & Aug. 2023 & 45.7\% &  \\
\multicolumn{1}{l|}{} & Mistral & 7B & Dec & Sep. 2023 & - &  \\
\multicolumn{1}{l|}{} & DSCoder & 6.7B & Dec & Dec. 2023 & 80.2\% &  \\
\multicolumn{1}{l|}{} & StarCoder2 & 7B & Dec & Feb. 2024 & 34.1\% &  \\
\multicolumn{1}{l|}{} & CodeQwen & 7B & Dec & Apr. 2024 & 50.8\% &  \\
\multicolumn{1}{l|}{} & LLama3 & 8B & Dec & Apr. 2024 & 62.2\% &  \\
\multicolumn{1}{l|}{} & Phi3 & 3.8B & Dec & May 2024 & 58.5\% &  \\ \bottomrule
\end{tabular}
}
\label{tab:studied_llm}
\vspace{-0.5cm}
\end{table}

\vspace{-0.3cm}
\subsection{LLM Adaptation Techniques}
We employ two typical techniques for adapting LLMs: prompt-based methods and fine-tuning-based methods. The last column of Table~\ref{tab:studied_llm} summarizes the corresponding LLM adaptation techniques used for each of the studied LLMs.

\vspace{-0.2cm}
\subsubsection{Prompt-based Methods}
Prompts act as guides for LLMs, formatting the input data and providing a natural language description of the task~\cite{gpt3}. Prompt-based methods mainly leverage the pre-trained knowledge of LLMs to perform the downstream task without the need for extensive task-specific training.
The prompt format designed for our vulnerability detection task comprises three components, organized sequentially: 1) Task Description: a natural language description of the vulnerability detection task, i.e.,  \textit{``If the following code snippet has any vulnerabilities, output Yes; otherwise, output No''};
2) Formatted Input: the input code is enclosed within two markers \textit{``// Code Start''} and \textit{``// Code End''};
3) Prediction Marker: a marker (\textit{``// Detection''}) instructing the model to start to provide its prediction.

We studied three popular prompt-based methods based on the aforementioned prompt designed for vulnerability detection:

\vspace{0.05cm}
\noindent
\textbf{Zero-shot Prompting}: 
In this approach, the input source code is formatted based on the prompt above and then being inputted into LLMs.
The LLMs then generate a response to detect whether the source code is vulnerable or not. It is named ``zero-shot'' because no labeled data is used in this approach.

\vspace{0.05cm}
\noindent
\textbf{Chain-of-thought (CoT) Prompting}: 
We utilized the Chain-of-thought technique proposed by Kojima et al.~\cite{cot} for improved reasoning, which involves adding \textit{``Let's think step by step''} to the original prompt. Specifically, we add the \textit{``Let's think step by step''} after the task description of the prompt above.

\vspace{0.05cm}
\noindent
\textbf{Few-shot Prompting}: 
In this approach, we provided a few examples of inputs and ground truth label pairs for LLMs. Those input-label pairs are concatenated after the original prompt. 
Specifically, we randomly select 2 examples from the training set, one vulnerable function, and one clean function.
Those examples can help the model to better understand the target task and reply in the same format as the example responses.

\vspace{-0.2cm}
\subsubsection{Fine-tuning-based Methods}
Fine-tuning is another widely used adaptation technique for LLMs, which often unlocks their full potential for specific tasks~\cite{wei2021finetuned}.
Fine-tuning involves training an LLM on task-specific data, enabling it to acquire domain knowledge and generate more relevant and meaningful output. In this study, we considered two different fine-tuning approaches based on the size of the LLM models:

\vspace{0.05cm}
\noindent
\textbf{Full Fine-Tuning on Lightweight LLMs.}
For the four lightweight LLMs with fewer than 1 billion parameters, i.e., CodeBERT, GraphCodeBERT, CodeT5, and UniXcoder, we utilized full parameter fine-tuning. This approach updates all model parameters to adapt the LLMs to the training data. 
Full parameter fine-tuning is the standard approach for these lightweight LLMs across various code-related tasks such as code search and  defect prediction~\cite{codebert,graphcodebert,codet5,unixcoder}.

\vspace{0.05cm}
\noindent
\textbf{Parameter-Efficient Fine-Tuning (PEFT) on Large LLMs.} 
For the eight large LLMs with more than 1 billion parameters such as DeepSeekCoder (DSCoder), we utilized the parameter-efficient fine-tuning technique to adapt them to the training data. This is because full-parameter fine-tuning comes at a significant computational cost when the LLM contains billions of parameters, exceeding our current computational resources~\cite{weyssow2023exploring}. 
Additionally, recent studies~\cite{weyssow2023exploring,qlora} suggest that parameter-efficient tuning techniques can achieve comparable or even better performance on LLMs exceeding 1 billion parameters.
Technically, parameter-efficient tuning techniques selectively update a subset of the model's parameters, significantly reducing the computational resources required~\cite{meng2024pissa,lora,qlora}.
In this work, we employ a popular and widely used parameter-efficient tuning technique called LoRA (Low-Rank Adaptation)~\cite{lora}, which has been widely used in many existing studies~\cite{liu2022few,ding2022delta,treviso2023efficient,weyssow2023exploring}.
LoRA allows the LLM to adapt to the target task while preserving its pre-trained knowledge and reducing the computational requirements compared to full-parameter fine-tuning~\cite{weyssow2023exploring}.

\vspace{-0.2cm}
\section{Experimental Setup}
\subsection{Vulnerability Detection Scenarios}
\label{DetectionScenarios}
Aligned with the previous work~\cite{C_CVE}, we assess the vulnerability detection capabilities of SAST tools and LLMs across two distinct scenarios. Specifically, we investigated:
\begin{itemize}[leftmargin=*,noitemsep,nolistsep]
    \item[-] \textbf{Scenario 1 (S1)}: A vulnerability is considered detected if the SAST tool or LLM identifies at least one vulnerable function within the vulnerable repository.
    \item[-] \textbf{Scenario 2 (S2)}: A vulnerability is only considered detected if the SAST tool or LLM identifies all vulnerable functions present in the vulnerable repository.
\end{itemize}
These scenarios underscore the significance of tools capable of pinpointing vulnerabilities within repositories precisely. By accurately identifying the vulnerable functions associated with potential security weaknesses, such tools streamline the manual search and remediation process for developers, enabling them to concentrate their efforts effectively on the identified vulnerable functions.

\vspace{-0.2cm}
\subsection{Evaluation Metrics}
\label{EvaluationMetrics}
We follow the previous work ~\cite{C_CVE} to employ the following metrics:
$$Vuln. \text{ } Detection \text{ } Ratio = \frac{\# Detected \text{ } Vuln.}{\# All \text{ } Vuln. \text{ } in \text{ } Benchmark} $$
$$Marked \text{ } Function \text{ } Ratio = \frac{\# Marked \text{ } Function}{\# All \text{ } Functions \text{ } in \text{ } Benchmark} $$
The first metric (i.e., the vulnerability detection ratio) computes the proportion of detected vulnerabilities present in the benchmark. 
A high vulnerability detection ratio signifies superior performance in vulnerability detection.
The second metric (i.e., the marked function ratio) approximates the false positive rate~\cite{C_CVE} because, on average, only 0.008\%--0.18\% of functions are vulnerable in the repositories, meaning that the majority of marked functions are non-vulnerable.
A low marked function ratio is desirable, as it reduces the number of functions developers need to inspect manually.

We refrain from employing the commonly used evaluation metric Precision in this study, aligning with previous research~\cite{C_CVE}. This decision stems from the inherent challenges in assessing real-world programs, which may conceal undiscovered vulnerabilities. Our knowledge is confined to vulnerabilities reported or identified by the community. Consequently, we lack precise information on the total number of positive samples (vulnerable functions), encompassing both known and unknown vulnerable functions.
Precision requires precise information on all positive samples, rendering its calculation unfeasible in our context. We also cannot compute the F1 score, as it is based on Precision.

\begin{table*}[t]
\centering
\setlength{\tabcolsep}{2mm}
\renewcommand{\arraystretch}{1}
\caption{
Experimental results of repo-level vulnerability detection. 
``S1 D'' refers to the vulnerability detection ratio in Scenario 1;
``S2 D'' refers to the vulnerability detection ratio in Scenario 2;
``Marked'' refers to the marked function ratio.
}
\vspace{-0.3cm}
\resizebox{0.95\textwidth}{!}{
\begin{tabular}{lclccccccccc}
\hline
\multicolumn{3}{c}{\textbf{Benchmarks}}                                                                                                                                              & \multicolumn{3}{c}{\textbf{Java}}                                               & \multicolumn{3}{c}{\textbf{C}}                                                  & \multicolumn{3}{c}{\textbf{Python}}                           \\ \hline
\multicolumn{2}{c}{\textbf{Categories}}                                                                                                                           & \textbf{Method}  & \textbf{S1 D$\uparrow$} & \textbf{S2 D$\uparrow$} & \textbf{Marked$\downarrow$}                   & \textbf{S1 D$\uparrow$} & \textbf{S2 D$\uparrow$} & \textbf{Marked$\downarrow$}                   & \textbf{S1 D$\uparrow$} & \textbf{S2 D$\uparrow$} & \textbf{Marked$\downarrow$} \\ \hline
\multicolumn{1}{l|}{\cellcolor[HTML]{EFEFEF}}                                 &                                                                                   & CodeQL           & 0.0                  & 0.0                  & \multicolumn{1}{c|}{0.2}          & -                    & -                    & \multicolumn{1}{c|}{-}            & -                    & -                    & -               \\
\multicolumn{1}{l|}{\cellcolor[HTML]{EFEFEF}}                                 &                                                                                   & Contrast         & 0.0                  & 0.0                  & \multicolumn{1}{c|}{0.0}          & -                    & -                    & \multicolumn{1}{c|}{-}            & -                    & -                    & -               \\
\multicolumn{1}{l|}{\cellcolor[HTML]{EFEFEF}}                                 &                                                                                   & Horusec          & {\ul 37.5}           & 0.0                  & \multicolumn{1}{c|} {\cellcolor{lightgreen}\textbf{1.7}} & -                    & \multicolumn{1}{c}-                    & \multicolumn{1}{c|}{-}            & -                    & -                    & -               \\
\multicolumn{1}{l|}{\cellcolor[HTML]{EFEFEF}}                                 &                                                                                   & Insider          & 0.0                  & 0.0                  &  \multicolumn{1}  {c|} {1.4}          & -                    & -                    & \multicolumn{1}{c|}{-}            & -                    & -                    & -               \\
\multicolumn{1}{l|}{\cellcolor[HTML]{EFEFEF}}                                 &                                                                                   & SpotBugs         & 0.0                  & 0.0                  & \multicolumn{1}{c|}{0.0}          & -                    & -                    & \multicolumn{1}{c|}{-}            & -                    & -                    & -               \\
\multicolumn{1}{l|}{\cellcolor[HTML]{EFEFEF}}                                 &                                                                                   & Semgrep          & 25.0                 & {\ul 0.0}           &  \multicolumn{1}{c|}{\cellcolor{lightgreen!20} \underline{2.1}}          & -                    & \multicolumn{1}{c}-                    & \multicolumn{1}{c|}{-}            & -                    & -                    & -               \\
\multicolumn{1}{l|}{\cellcolor[HTML]{EFEFEF}}                                 & \multirow{-7}{*}{\textbf{Java}}                                                   & SonarQube        & 0.0                  & 0.0                  & \multicolumn{1}{c|} {0.6}          & -                    & -                    & \multicolumn{1}{c|}{-}            & -                    & -                    & -               \\ \cline{2-12} 
\multicolumn{1}{l|}{\cellcolor[HTML]{EFEFEF}}                                 &                                                                                   & Flawfinder       & -                    & -                    & \multicolumn{1}{c|}{-}            & 0.0                  & 0.0                  & \multicolumn{1}{c|}{1.9}          & -                    & -                    & -               \\
\multicolumn{1}{l|}{\cellcolor[HTML]{EFEFEF}}                                 &                                                                                   & Cppcheck         & -                    & -                    & \multicolumn{1}{c|}{-}            & 0.0                  & 0.0                  & \multicolumn{1}{c|}{0.3}          & -                    & -                    & -               \\
\multicolumn{1}{l|}{\cellcolor[HTML]{EFEFEF}}                                 &                                                                                   & Infer            & -                    & -                    & \multicolumn{1}{c|}{-}            & 11.1                 & 11.1                 & \multicolumn{1}{c|}{\cellcolor{lightgreen} \textbf{0.9}} & -                    & -                    & -               \\
\multicolumn{1}{l|}{\cellcolor[HTML]{EFEFEF}}                                 &                                                                                   & CodeChecker      & -                    & -                    & \multicolumn{1}{c|}{-}            & 0.0                  & 0.0                  & \multicolumn{1}{c|}{0.3}          & -                    & -                    & -               \\
\multicolumn{1}{l|}{\cellcolor[HTML]{EFEFEF}}                                 & \multirow{-5}{*}{\textbf{C}}                                                      & CodeQL           & -                    & -                    & \multicolumn{1}{c|}{-}            & {\ul 44.4}           & {\ul 33.3}           & \multicolumn{1}{c|}{5.2}          & -                    & -                    & -               \\ \cline{2-12} 
\multicolumn{1}{l|}{\cellcolor[HTML]{EFEFEF}}                                 &                                                                                   & Badnit           & -                    & -                    & \multicolumn{1}{c|}{-}            & -                    & -                    & \multicolumn{1}{c|}{-}            & 20.0                 & {\ul 20.0}           & \cellcolor{lightgreen!20} \underline{2.4}             \\
\multicolumn{1}{l|}{\cellcolor[HTML]{EFEFEF}}                                 &                                                                                   & Dlint            & -                    & -                    & \multicolumn{1}{c|}{-}            & -                    & -                    & \multicolumn{1}{c|}{-}            & 20.0                 &  {\ul 20.0}           & \cellcolor{lightgreen} \textbf{0.6}    \\
\multicolumn{1}{l|}{\cellcolor[HTML]{EFEFEF}}                                 &                                                                                   & DevSkim          & -                    & -                    & \multicolumn{1}{c|}{-}            & -                    & -                    & \multicolumn{1}{c|}{-}            & {\ul 30.0}           & {\ul 20.0}           & 4.5             \\
\multicolumn{1}{l|}{\cellcolor[HTML]{EFEFEF}}                                 &                                                                                   & CodeQL           & -                    & -                    & \multicolumn{1}{c|}{-}            & -                    & -                    & \multicolumn{1}{c|}{-}            & 0.0                  & 0.0                  & 0.4             \\
\multicolumn{1}{l|}{\cellcolor[HTML]{EFEFEF}}                                 &                                                                                   & Graudit          & -                    & -                    & \multicolumn{1}{c|}{-}            & -                    & -                    & \multicolumn{1}{c|}{-}            & 0.0                  & 0.0                  & 0.04            \\
\multicolumn{1}{l|}{\multirow{-18}{*}{\cellcolor[HTML]{EFEFEF}\textbf{SAST}}} & \multirow{-6}{*}{\textbf{Python}}                                                 & Semgrep          & -                    & -                    & \multicolumn{1}{c|}{-}            & -                    & -                    & \multicolumn{1}{c|}{-}            & 10.0                 & 10.0                 & \cellcolor{lightgreen} \textbf{0.6}    \\ \hline
\multicolumn{1}{l|}{\cellcolor[HTML]{ECF4FF}}                                 &                                                                                   & CodeBERT         & \cellcolor{lightgreen!20} \underline{87.5}                 & 50.0                 & \multicolumn{1}{c|}{35.4}         & \cellcolor{lightgreen} \textbf{100.0}       & \cellcolor{lightgreen} \textbf{88.9}        & \multicolumn{1}{c|}{36.2}         & \cellcolor{lightgreen!20} \underline{80.0}                 & 60.0                 & 41.2            \\
\multicolumn{1}{l|}{\cellcolor[HTML]{ECF4FF}}                                 &                                                                                   & GraphCodeBERT    & 75.0                 & 50.0                 & \multicolumn{1}{c|}{30.1}         & \cellcolor{lightgreen} \textbf{100.0}       & \cellcolor{lightgreen} \textbf{88.9}        & \multicolumn{1}{c|}{44.6}         & \cellcolor{lightgreen!20} \underline{80.0}                 & \cellcolor{lightgreen!20} \underline{70.0}                 & 50.0            \\
\multicolumn{1}{l|}{\cellcolor[HTML]{ECF4FF}}                                 &                                                                                   & CodeT5           & \cellcolor{lightgreen!20} \underline{87.5}                 & \cellcolor{lightgreen!20} \underline{62.5}                 & \multicolumn{1}{c|}{24.3}         & \cellcolor{lightgreen!20} \underline{88.9}                 & \cellcolor{lightgreen!20} \underline{77.8}                 & \multicolumn{1}{c|}{19.8}         & \cellcolor{lightgreen} \textbf{90.0}        & \cellcolor{lightgreen} \textbf{90.0}        & 50.5            \\
\multicolumn{1}{l|}{\cellcolor[HTML]{ECF4FF}}                                 & \multirow{-4}{*}{\textbf{\begin{tabular}[c]{@{}c@{}}Full \\ Tuning\end{tabular}}} & Unixcoder        & 75.0                 & \cellcolor{lightgreen!20} \underline{62.5}                 & \multicolumn{1}{c|}{21.4}         & \cellcolor{lightgreen} \textbf{100.0}                & \cellcolor{lightgreen} \textbf{88.9}                 & \multicolumn{1}{c|}{25.6}         & 70.0                 & 50.0                 & 42.0            \\ \cline{2-12} 
\multicolumn{1}{l|}{\cellcolor[HTML]{ECF4FF}}                                 &                                                                                   & CodeLlama     & 75.0                 & 37.5                 & \multicolumn{1}{c|}{{\ul 8.3}}    & 44.4                 & 44.4                 & \multicolumn{1}{c|}{7.4}          & \cellcolor{lightgreen!20} \underline{80.0}                 & 40.0                 & 39.5            \\
\multicolumn{1}{l|}{\cellcolor[HTML]{ECF4FF}}                                 &                                                                                   & Llama3        & \cellcolor{lightgreen}  \textbf{100.0}       & \cellcolor{lightgreen}  \textbf{87.5}        & \multicolumn{1}{c|}{21.8}         & 22.2                 & 22.2                 &  \multicolumn{1}{c|}{\cellcolor{lightgreen!20} {\underline{\ul 3.5}}}    & \cellcolor{lightgreen!20} \underline{80.0}                 & 60.0                 & 29.8            \\
\multicolumn{1}{l|}{\cellcolor[HTML]{ECF4FF}}                                 &                                                                                   & CodeQwen      & 75.0                 & 37.5                 & \multicolumn{1}{c|}{14.8}         & 22.2                 & 22.2                 & \multicolumn{1}{c|}{12.2}         & \cellcolor{lightgreen!20} \underline{80.0}                 & 60.0                 & 35.8            \\
\multicolumn{1}{l|}{\cellcolor[HTML]{ECF4FF}}                                 &                                                                                   & DeepSeek-Coder & 62.5                 & \cellcolor{lightgreen!20} \underline{62.5}                 & \multicolumn{1}{c|}{15.1}         & 11.1                 & 11.1                 & \multicolumn{1}{c|}{10.6}         & 50.0                 & 30.0                 & 17.3            \\
\multicolumn{1}{l|}{\cellcolor[HTML]{ECF4FF}}                                 &                                                                                   & Mistral       & 62.5                 & 50.0                 & \multicolumn{1}{c|}{21.8}         & 11.1                 & 11.1                 & \multicolumn{1}{c|}{6.5}          & 50.0                 & 10.0                 & 18.9            \\
\multicolumn{1}{l|}{\cellcolor[HTML]{ECF4FF}}                                 &                                                                                   & Phi3        & 50.0                 & 50.0                 & \multicolumn{1}{c|}{20.4}         & 33.3                 & 33.3                 & \multicolumn{1}{c|}{10.3}         & 50.0                 & 20.0                 & 20.8            \\
\multicolumn{1}{l|}{\cellcolor[HTML]{ECF4FF}}                                 &                                                                                   & Starcoder     & 75.0                 & 37.5                 & \multicolumn{1}{c|}{15.6}         & 44.4                 & 44.4                 & \multicolumn{1}{c|}{12.0}         & 50.0                 & 30.0                 & 26.5            \\
\multicolumn{1}{l|}{\cellcolor[HTML]{ECF4FF}}                                 & \multirow{-8}{*}{\textbf{PEFT}}                                                   & Starcoder2    & \cellcolor{lightgreen!20} \underline{87.5}                 & 50.0                 & \multicolumn{1}{c|}{37.5}         & 0.0                  & 0.0                  & \multicolumn{1}{c|}{0.3}          & 40.0                 & 40.0                 & 25.4            \\ \cline{2-12} 
\multicolumn{1}{l|}{\cellcolor[HTML]{ECF4FF}}                                 &                                                                                   & CodeLlama     & \cellcolor{lightgreen!20} \underline{87.5}                 & 50.0                 & \multicolumn{1}{c|}{66.9}         & 44.4                 & 44.4                 & \multicolumn{1}{c|}{47.0}         & 70.0                 & 40.0                 & 51.2            \\
\multicolumn{1}{l|}{\cellcolor[HTML]{ECF4FF}}                                 &                                                                                   & Llama3        & \cellcolor{lightgreen!20} \underline{87.5}                 & \cellcolor{lightgreen!20} \underline{62.5}                 & \multicolumn{1}{c|}{47.3}         & 66.7                 & 55.6                 & \multicolumn{1}{c|}{49.4}         & 50.0                 & 20.0                 & 40.1            \\
\multicolumn{1}{l|}{\cellcolor[HTML]{ECF4FF}}                                 &                                                                                   & CodeQwen      & 50.0                 & 37.5                 & \multicolumn{1}{c|}{60.1}         & 33.3                 & 33.3                 & \multicolumn{1}{c|}{53.7}         & 70.0                 & 30.0                 & 51.3            \\
\multicolumn{1}{l|}{\cellcolor[HTML]{ECF4FF}}                                 &                                                                                   & DeepSeek-Coder & 37.5                 & 12.5                 & \multicolumn{1}{c|}{35.4}         & 11.0                 & 0.0                  & \multicolumn{1}{c|}{24.3}         & 70.0                 & 40.0                 & 37.1            \\
\multicolumn{1}{l|}{\cellcolor[HTML]{ECF4FF}}                                 &                                                                                   & Mistral       & 50.0                 & 25.0                 & \multicolumn{1}{c|}{64.6}         & 55.6                 & 55.6                 & \multicolumn{1}{c|}{60.5}         & \cellcolor{lightgreen!20} \underline{80.0}                 & 50.0                 & 52.4            \\
\multicolumn{1}{l|}{\cellcolor[HTML]{ECF4FF}}                                 &                                                                                   & Phi3        & 62.5                 & 25.0                 & \multicolumn{1}{c|}{41.9}         & 77.8                 & 66.7                 & \multicolumn{1}{c|}{66.0}         & 20.0                 & 0.0                  & 30.8            \\
\multicolumn{1}{l|}{\cellcolor[HTML]{ECF4FF}}                                 &                                                                                   & Starcoder     & 62.5                 & 37.5                 & \multicolumn{1}{c|}{73.4}         & 66.7                 & 55.6                 & \multicolumn{1}{c|}{54.1}         & 50.0                 & 20.0                 & 55.9            \\
\multicolumn{1}{l|}{\cellcolor[HTML]{ECF4FF}}                                 & \multirow{-8}{*}{\textbf{Zero-shot}}                                              & Starcoder2    & 37.5                 & 12.5                 & \multicolumn{1}{c|}{29.3}         & 44.4                 & 44.4                 & \multicolumn{1}{c|}{25.6}         & 20.0                 & 0.0                  & 15.2            \\ \cline{2-12} 
\multicolumn{1}{l|}{\cellcolor[HTML]{ECF4FF}}                                 &                                                                                   & CodeLlama     & 75.0                 & 50.0                 & \multicolumn{1}{c|}{70.4}         & 55.6                 & 44.4                 & \multicolumn{1}{c|}{50.0}         & 60.0                 & 20.0                 & 57.4            \\
\multicolumn{1}{l|}{\cellcolor[HTML]{ECF4FF}}                                 &                                                                                   & Llama3        & 75.0                 & 37.5                 & \multicolumn{1}{c|}{47.4}         & 44.4                 & 33.3                 & \multicolumn{1}{c|}{46.9}         & \cellcolor{lightgreen!20} \underline{80.0}                 & 40.0                 & 42.2            \\
\multicolumn{1}{l|}{\cellcolor[HTML]{ECF4FF}}                                 &                                                                                   & CodeQwen      & 75.0                 & 37.5                 & \multicolumn{1}{c|}{66.4}         & 44.4                 & 33.3                 & \multicolumn{1}{c|}{53.3}         & \cellcolor{lightgreen!20} \underline{80.0}                 & 60.0                 & 51.0            \\
\multicolumn{1}{l|}{\cellcolor[HTML]{ECF4FF}}                                 &                                                                                   & DeepSeek-Coder & 25.0                 & 25.0                 & \multicolumn{1}{c|}{29.2}         & 0.0                  & 0.0                  & \multicolumn{1}{c|}{18.0}         & 30.0                 & 0.0                  & 29.6            \\
\multicolumn{1}{l|}{\cellcolor[HTML]{ECF4FF}}                                 &                                                                                   & Mistral       & 62.5                 & 37.5                 & \multicolumn{1}{c|}{69.2}         & 77.8                 & 66.7                 & \multicolumn{1}{c|}{64.8}         & 70.0                 & 50.0                 & 57.7            \\
\multicolumn{1}{l|}{\cellcolor[HTML]{ECF4FF}}                                 &                                                                                   & Phi3        & 12.5                 & 0.0                  & \multicolumn{1}{c|}{44.9}         & 66.7                 & 66.7                 & \multicolumn{1}{c|}{64.6}         & 50.0                 & 20.0                 & 32.6            \\
\multicolumn{1}{l|}{\cellcolor[HTML]{ECF4FF}}                                 &                                                                                   & Starcoder     & \cellcolor{lightgreen!20} \underline{87.5}                 & 50.0                 & \multicolumn{1}{c|}{70.6}         & 55.6                 & 44.4                 & \multicolumn{1}{c|}{50.8}         & 60.0                 & 40.0                 & 53.7            \\
\multicolumn{1}{l|}{\cellcolor[HTML]{ECF4FF}}                                 & \multirow{-8}{*}{\textbf{CoT}}                                                    & Starcoder2    & 75.0                 & 50.0                 & \multicolumn{1}{c|}{41.9}         & 55.6                 & 44.4                 & \multicolumn{1}{c|}{28.3}         & 40.0                 & 20.0                 & 22.2            \\ \cline{2-12} 
\multicolumn{1}{l|}{\cellcolor[HTML]{ECF4FF}}                                 &                                                                                   & CodeLlama     & 62.5                 & 37.5                 & \multicolumn{1}{c|}{74.7}         & 66.7                 & 55.6                 & \multicolumn{1}{c|}{59.2}         & \cellcolor{lightgreen} \textbf{90.0}        & \cellcolor{lightgreen!20} \underline{70.0}                 & 64.3            \\
\multicolumn{1}{l|}{\cellcolor[HTML]{ECF4FF}}                                 &                                                                                   & Llama3        & 62.5                 & 25.0                 & \multicolumn{1}{c|}{29.8}         & 44.4                 & 44.4                 & \multicolumn{1}{c|}{43.1}         & 70.0                 & 60.0                 & 35.9            \\
\multicolumn{1}{l|}{\cellcolor[HTML]{ECF4FF}}                                 &                                                                                   & CodeQwen      & 50.0                 & 25.0                 & \multicolumn{1}{c|}{61.1}         & 55.6                 & 44.1                 & \multicolumn{1}{c|}{47.7}         & 70.0                 & 50.0                 & 45.4            \\
\multicolumn{1}{l|}{\cellcolor[HTML]{ECF4FF}}                                 &                                                                                   & DeepSeek-Coder & 12.5                 & 12.5                 & \multicolumn{1}{c|}{26.0}         & 0.0                  & 0.0                  & \multicolumn{1}{c|}{5.4}          & 30.0                 & 10.0                 & {\ul 14.9}      \\
\multicolumn{1}{l|}{\cellcolor[HTML]{ECF4FF}}                                 &                                                                                   & Mistral       & 25.0                 & 12.5                 & \multicolumn{1}{c|}{22.0}         & 33.3                 & 22.2                 & \multicolumn{1}{c|}{28.7}         & 50.0                 & 20.0                 & 22.4            \\
\multicolumn{1}{l|}{\cellcolor[HTML]{ECF4FF}}                                 &                                                                                   & Phi3        & 75.0                 & 37.5                 & \multicolumn{1}{c|}{59.2}         & 44.4                 & 44.4                 & \multicolumn{1}{c|}{58.3}         & 40.0                 & 10.0                 & 28.9            \\
\multicolumn{1}{l|}{\cellcolor[HTML]{ECF4FF}}                                 &                                                                                   & Starcoder     & 75.0                 & 50.0                 & \multicolumn{1}{c|}{66.8}         & 55.6                 & 55.6                 & \multicolumn{1}{c|}{45.3}         & \cellcolor{lightgreen!20} \underline{80.0}                 & 50.0                 & 46.7            \\
\multicolumn{1}{l|}{\multirow{-36}{*}{\cellcolor[HTML]{ECF4FF}\textbf{LLM}}}  & \multirow{-8}{*}{\textbf{Few-shot}}                                               & Starcoder2    & 62.5                 & 37.5                 & \multicolumn{1}{c|}{77.4}         & 55.6                 & 55.6                 & \multicolumn{1}{c|}{70.1}         & 70.0                 & 60.0                 & 73.4            \\ \hline
\end{tabular}
}
\label{table:RQ1}
\vspace{-0.3cm}
\end{table*}

\vspace{-0.2cm}
\subsection{Implementation Details}
\textbf{Data Split.}
Following the previous work~\cite{devign,codebert,graphcodebert}, we randomly split each vulnerability dataset into disjoint training, validation, and test sets in a ratio of 8:1:1. The training and validation sets are used to fine-tune/prompt LLMs. Both SAST and LLMs are evaluated on the same test sets for fair comparison.

\vspace{0.05cm}
\noindent\textbf{Implementations.}
For prompt-based methods, any required labeled data (e.g., for few-shot prompting) is randomly sampled from the training set. In contrast, fine-tuning-based methods employ a different approach. Firstly, we identify and extract all vulnerable functions from the repositories in the training set. Subsequently, we randomly select an equal number of non-vulnerable functions from the repositories in the training set. By combining these two groups of functions, we construct a balanced training set for fine-tuning LLMs. 
This balancing process is essential for effective model training when dealing with highly imbalanced datasets~\cite{hoang2019deepjit,DBLP:conf/icse/YangWLW23,xu2021empirical}. Without the balancing, models would be overly biased towards learning the features of non-vulnerable functions due to their dominance in the datasets, resulting in sub-optimal performance in detecting vulnerable samples~\cite{zhou2023devil}.
Similarly, we also construct a balanced validation set for model validation.
Additionally, for LLMs, we develop a generation pipeline in Python, utilizing PyTorch implementations of studied LLMs. We leverage the Hugging Face library~\cite{HF} to load the model weights and generate outputs.
We fine-tune all LLMs for 20 epochs, with a batch size of 8, and select the best model snapshot based on the validation performance. For the parameter-efficient tuning technique LoRA, we set its parameters as $r=8$ and $\alpha=16$. All the experiments are conducted on an Ubuntu 22.04 server equipped with A5000 GPU.

\vspace{-0.2cm}
\subsection{Research Questions}
Our work aims to answer three research questions (RQs). 
\begin{itemize} [leftmargin=*]
\item \textbf{RQ1: How effective are SAST tools and LLMs?} 
In RQ1, we evaluate a comprehensive list of 15 SAST tools and 12 LLMs in their ability to detect vulnerabilities in software repositories.

\item \textbf{RQ2: Which one is more effective for detecting vulnerabilities, SAST tools or LLMs?} 
In RQ2, we compare the effectiveness of SAST tools against LLMs, aiming to determine which approach performs better in detecting vulnerabilities across different programming languages.

\item \textbf{RQ3: To what extent can combining multiple SAST tools or LLMs improve vulnerability detection?} 
In RQ3, we investigate the potential enhancements in vulnerability detection achieved by integrating multiple SAST tools and LLMs.

\end{itemize}

\vspace{-0.2cm}
\section{Experimental Results}

\subsection{RQ1: Effectiveness of SAST Tools and LLMs}

\begin{figure*} [t]
  \begin{minipage}{0.33\textwidth}
    \centering
    \includegraphics[width=\textwidth]{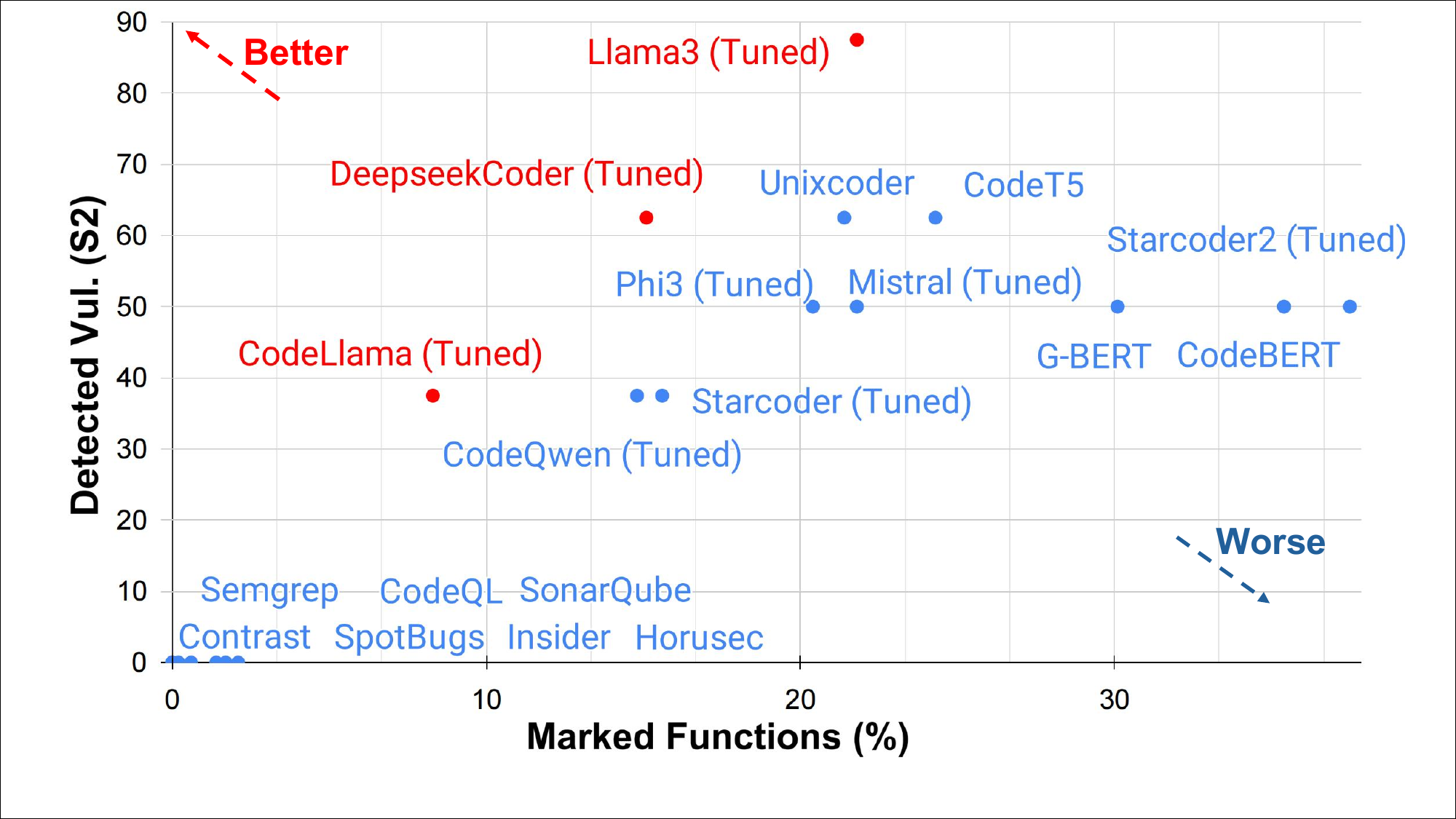}
    \vspace{-0.5cm}
    \subcaption{Java Benchmark}
  \end{minipage}
  \begin{minipage}{0.33\textwidth}
    \centering
    \includegraphics[width=\textwidth]{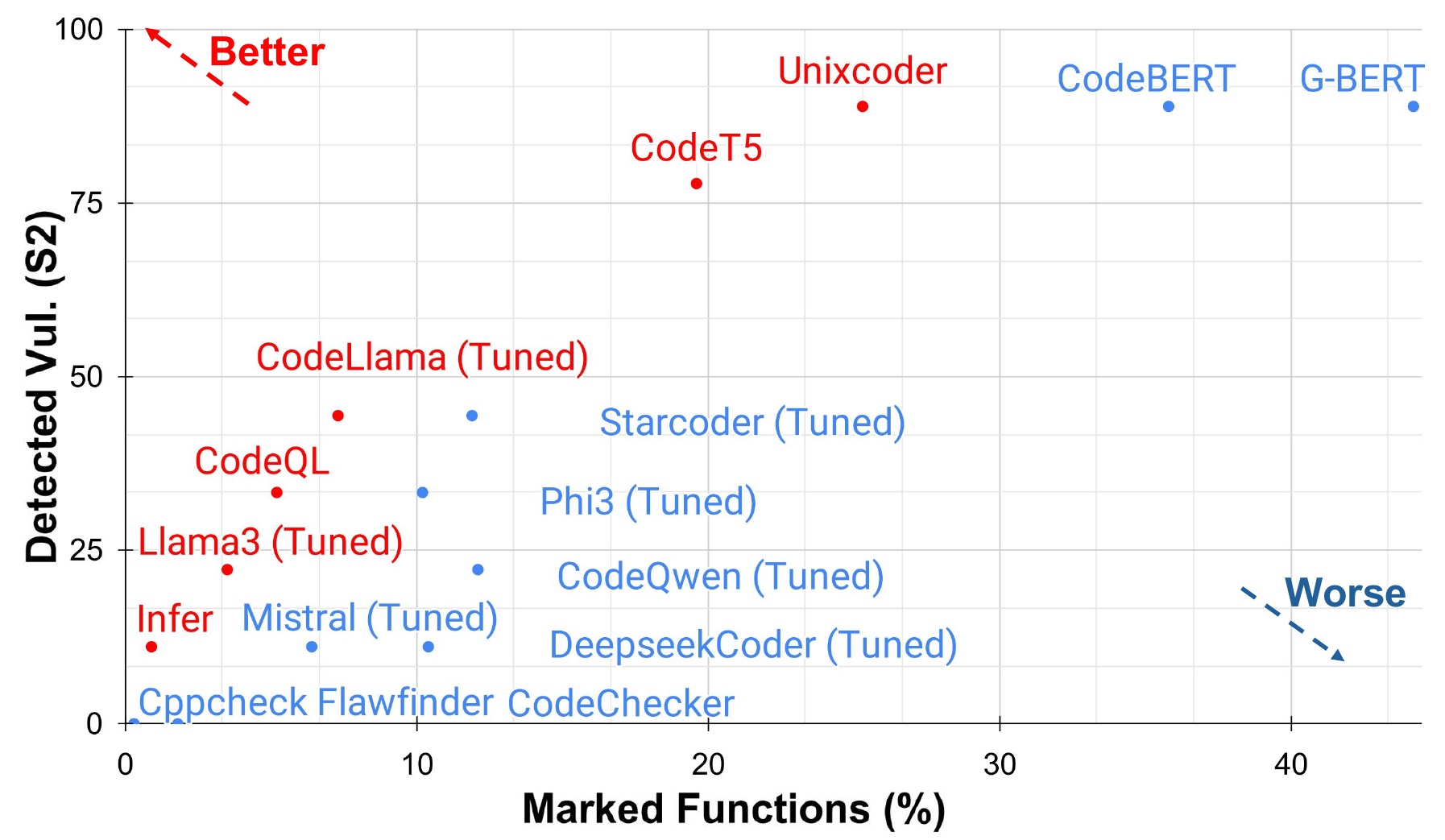}
    \vspace{-0.5cm}
    \subcaption{C Benchmark}
  \end{minipage}
  \begin{minipage}{0.33\textwidth}
    \centering
    \includegraphics[width=\textwidth]{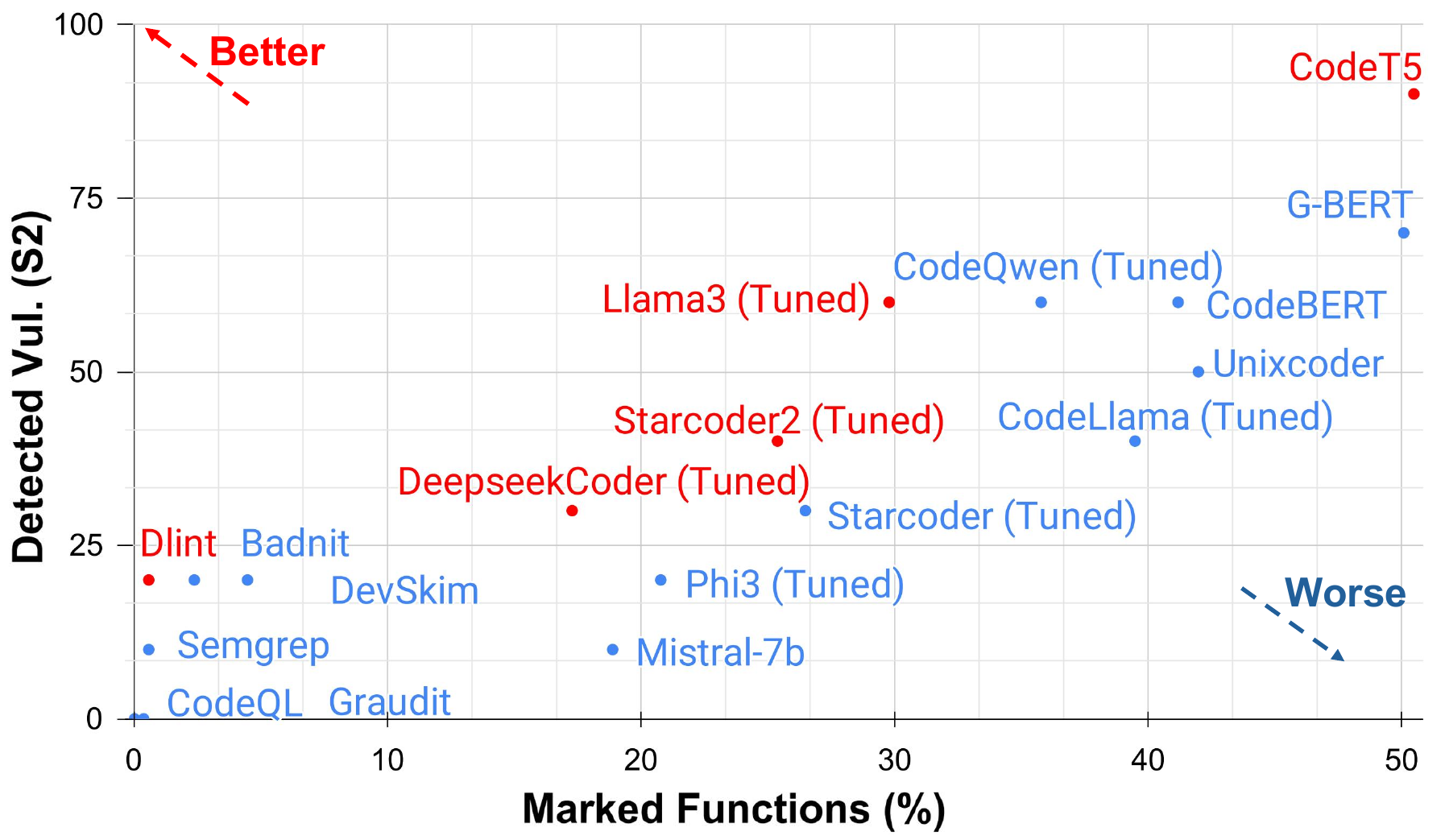}
    \vspace{-0.5cm}
    \subcaption{Python Benchmark}
  \end{minipage}
  \vspace{-0.4cm}
  \caption{The Relationship between the Vulnerability Detection Ratios (Scenario 2) and the Marked Function Ratios
  }
  \vspace{-0.3cm}
  \label{fig:tradeoff}
\end{figure*}

Table~\ref{table:RQ1} displays the experimental results, encompassing metrics such as ``S1 D,'' ``S2 D,'' and ``Marked.''
The ``S1 D'' metric refers to the ratio of detected vulnerabilities in Scenario 1, where detection is considered successful upon identifying any vulnerable function within the vulnerable repository. 
The ``S2 D'' metric represents the detection ratio in Scenario 2, where a vulnerability is considered as detected only when all vulnerable functions in the repository are identified.
The "Marked" metric indicates the percentage of functions labeled as vulnerable.
A higher detection ratio signifies superior performance in vulnerability detection, whereas a lower marked function ratio is preferred, as it reduces the number of functions developers need to check to locate the actual vulnerabilities.
In Table~\ref{table:RQ1}, bold numbers with green cells indicate the best performance, and underlined numbers with light green cells indicate the second-best performance. 
Please note that for the best marked function ratio, we consider the tool or model that achieves the lowest marked function ratio while still detecting at least one vulnerability. A tool or model that fails to detect any vulnerabilities, regardless of its marked function ratio, is not considered to have the best marked function ratio.

 \vspace{-0.1cm}
\subsubsection{Effectiveness of SAST Tools}
\textbf{SAST tools achieved low detection rates.} Many SAST tools, such as CodeQT, Contrast, Insider, SpotBugs, SonarQube for Java, FlawFinder, CppChecker, and CodeChecker for C, and CodeQL and Graudit for Python, failed to detect vulnerabilities.
Among SAST tools, Horusec performed the best for Java, achieving a 37.5\% detection rate in Scenario 1. For C, CodeQL exhibited the best performance, attaining detection rates of 44.4\% and 33.3\% in Scenario 1 and 2. For Python, DevSkim performed the best with detection rates of 30.0\% and 20.0\% in Scenario 1 and 2.
\textbf{The advantage of SAST tools is they marked a small percentage of functions as vulnerable.} Low marked function ratios could facilitate developers in identifying the actual vulnerability location by inspecting fewer marked functions.
SAST tools marked only 0.2\% to 5.2\% of all functions on average.

 \vspace{-0.1cm}
\subsubsection{Effectiveness of LLMs}
\textbf{LLMs achieved high detection rates.}
For Java, many LLMs exhibited detection rates ranging from over 50\% to as high as 100\% in Scenario 1, and from surpassing 37.5\% to as high as 87.5\% in Scenario 2.
Similarly, for C, LLMs achieved detection rates up to 100.0\% and 88.9\% in Scenario 1 and Scenario 2.
For Python, LLMs achieved detection rates up to 90.0\% in both Scenario 1 and Scenario 2.
\textbf{However, the drawback of LLMs is they mark a large portion of functions as vulnerable.}
This indicates that their high detection rates come at the cost of marking many functions in the repository as suspicious. Consequently, this makes it difficult for developers to pinpoint the actual vulnerability locations, as they need to inspect a large number of marked functions. Specifically, LLMs mark 8.3\% to 77.4\% of functions in Java, 3.5\% to 70.1\% in C, and 14.9\% to 73.4\% in Python.

\vspace{-0.1cm}
\subsubsection{Comparisons Among LLMs}
\textbf{Among different adaptations for large LLMs ($\geq$1B), fine-tuning achieves the highest detection ratios with the fewest marked vulnerable functions.} 
For instance, in Java, the fine-tuned large LLMs detect an average of 73.4\% and 51.6\% of vulnerabilities in S1 and S2 respectively, across different LLM models. This outperforms the prompt-based methods by 20.5\% to 73.7\% in terms of average detection ratios.
Meanwhile, the fine-tuned large LLMs mark only 19.4\% of functions as vulnerable on average, which is much less than the prompt-based methods, ranging from 52.1\% to 55.0\%.
Thus, fine-tuned large LLMs outperform prompted large LLMs.
\textbf{Fine-tuned lightweight LLMs detect more vulnerabilities but also mark more functions as vulnerable compared to fine-tuned large LLMs.} 
For example, in Java, fine-tuned lightweight LLMs outperform the fine-tuned large LLMs by 10.6\% and 9.1\% in S1 and S2, on average.
However, at the same time, fine-tuned lightweight LLMs mark 43.2\% more functions as vulnerable on average.

\vspace{0.1cm}
\noindent
\begin{tcolorbox} [boxrule=0.8pt,
                top=0.2pt,
                  bottom=0.2pt]
    \textbf{Answer to RQ1}: 
    SAST tools obtained lower vulnerability detection rates (up to 44.4\%) while also marking fewer functions as vulnerable (up to 5.2\%). In contrast, LLMs detected more vulnerabilities (up to 100\%) but also marked a higher proportion of functions as vulnerable (up to 77.4\%). 
\end{tcolorbox}

\vspace{-0.3cm}
\subsection{RQ2: SAST Tools vs. LLMs}

\vspace{-0.1cm}
\subsubsection{Comprehensive Comparison between SAST and LLMs}
The advantages and disadvantages of SAST tools and LLMs are evident: 1) SAST tools mark fewer functions but detect fewer vulnerabilities, while 2) LLMs detect more vulnerabilities but require marking substantially more functions. 
There seems to be a trade-off between detection ratios and marked function ratios: the more vulnerabilities detected, the greater the number of functions marked as potentially vulnerable. 
Thus, our aim here is to identify the optimal SAST tools and LLMs that reach a sweet spot in the trade-off between detection ratios and marked function ratios, optimizing both metrics for practical vulnerability detection scenarios.

To comprehensively understand the capabilities of SAST tools and LLMs, we plot the relationships between detection ratios (Scenario 2) and marked function ratios in Figure~\ref{fig:tradeoff}. The optimal tools/models are those capable of detecting more vulnerabilities with fewer marked functions, typically occupying the upper-left areas of Figure~\ref{fig:tradeoff}.
In Figure~\ref{fig:tradeoff}, we exclude prompt-based large LLMs because they usually appear in the bottom-right area (low detection ratios with high marked function ratios). 

We highlight the tools/models in red in the upper-right area of the figures, indicating those are more optimal when comprehensively considering both metrics.
\textbf{Our analysis reveals that the optimal approaches differ across programming languages.}
For instance, fine-tuned CodeLlama, fine-tuned DeepSeekCoder, and fine-tuned Llama3 emerge as the optimal choices for Java. Similarly, for C, the optimal models/tools include Infer, fine-tuned Llama3, CodeQL, fine-tuned CodeLlama, CodeT5, and Unixcoder. For Python, the optimal tools/models are Dlint, fine-tuned DeepSeekCoder, fine-tuned StarCoder2, fine-tuned Llama3, and CodeT5.
\textbf{The best choice will depend on the user's acceptance of the trade-off between detection ratios and marked function ratios.} 
For instance, for Java, if a user requires a model/tool with at most a 10\% marked function ratio, the fine-tuned CodeLlama would be the best choice.

To identify the best approach objectively, we also propose a ranking-based metric:
1) We rank each approach based on two criteria: the detection ratio in S2 and the marked function ratio.
2) We then sum the ranks of two criteria. The approach with the lowest total rank is considered the best.
According to this ranking-based metric, \textbf{fine-tuned DeepseekCoder, UniXcoder, and Llama3 are the best approaches for Java, C, and Python, respectively.}

\vspace{-0.2cm}
\subsubsection{Best vs. Worst Detected Vulnerabilities}
We studied the detection rates of vulnerabilities belonging to different vulnerability classes. Due to the page limit, we simplify the analysis by merging the results corresponding to the three programming languages together. Additionally, we consider the combined results of SAST tools: a vulnerability is considered detected if at least one SAST tool can detect it.
Similarly, we consider the combined results of the 2 optimal LLMs. 
Specifically, we focus on the 2 optimal models highlighted in Figure~\ref{fig:tradeoff}, which have the lowest marked function ratios.
For Java, the 2 optimal LLMs are fine-tuned CodeLlama and DeepSeekCoder. For C, the 2 optimal LLMs are fine-tuned Llama3 and CodeLlama. For Python, the optimal LLMs are fine-tuned DeepSeekCoder and StarCoder2. A vulnerability is considered detected if at least one of the 2 optimal LLMs can detect it.

\begin{figure}[t]
    \centering
    \includegraphics[width=0.49\textwidth]{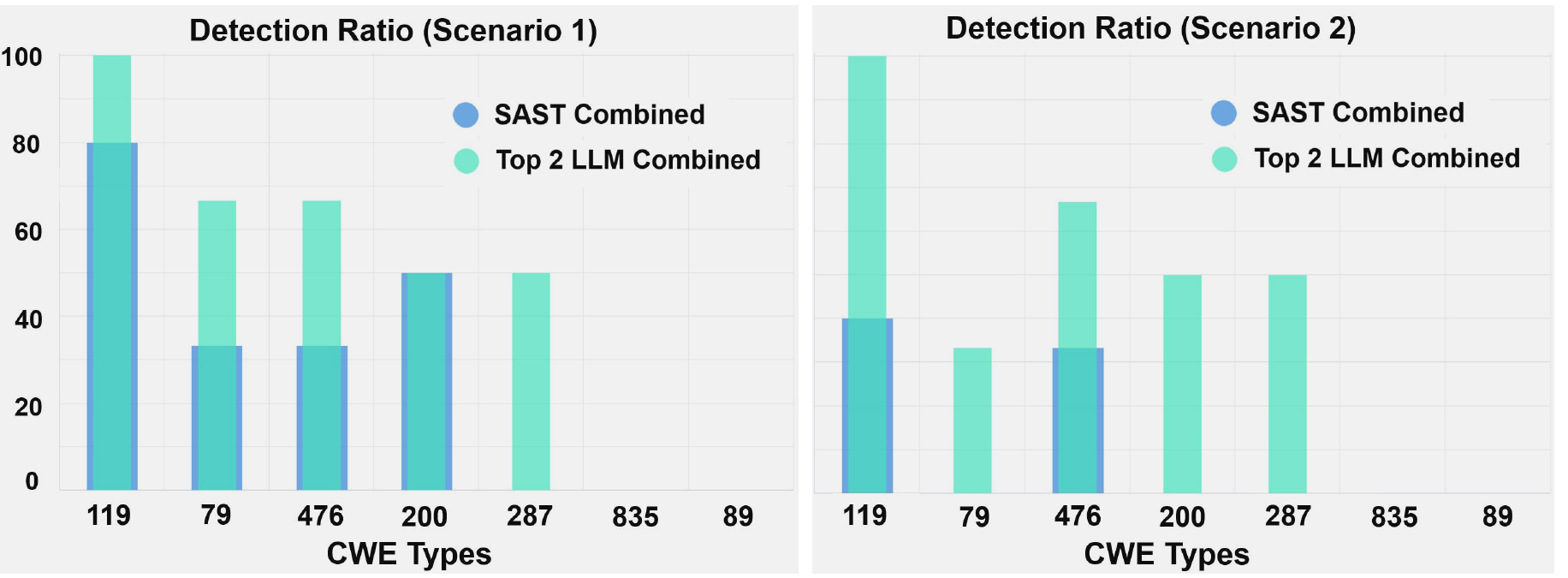}
    \vspace{-0.7cm}
    \caption{Vulnerability Detection Ratios in Different Classes}
  \label{fig:different_class}
  \vspace{-0.6cm}
\end{figure}

Figure~\ref{fig:different_class} illustrates the detection rates of vulnerability classes that are best and worst detected by SAST tools and the 2 optimal LLMs.
The most frequently detected vulnerability class in Scenarios 1 and 2 is CWE-119. Notably, the LLM group can achieve 100\% detection rates for CWE-119.
Additionally, CWE-476 is the second-best detected vulnerability class by both the SAST group and the LLM group. 
In contrast, CWE-835 and CWE-89 are the worst detected classes. Both the SAST tool group and the 2 optimal LLMs group failed to detect any vulnerabilities in these classes.

\vspace{0.1cm}
\noindent
\begin{tcolorbox} [boxrule=0.8pt,
                top=0.2pt,
                  bottom=0.2pt]
    \textbf{Answer to RQ2}: 
    The choice of the best approach depends on the programming language.
    Fine-tuned DeepseekCoder, UniXcoder, and Llama3 are the best approaches for Java, C, and Python, respectively.
\end{tcolorbox}

\vspace{-0.3cm}
\subsection{RQ3: Combining SAST Tools or LLMs}

 \subsubsection{Combining SAST Tools}
\begin{figure}[t]
    \centering
    \includegraphics[width=0.42\textwidth]{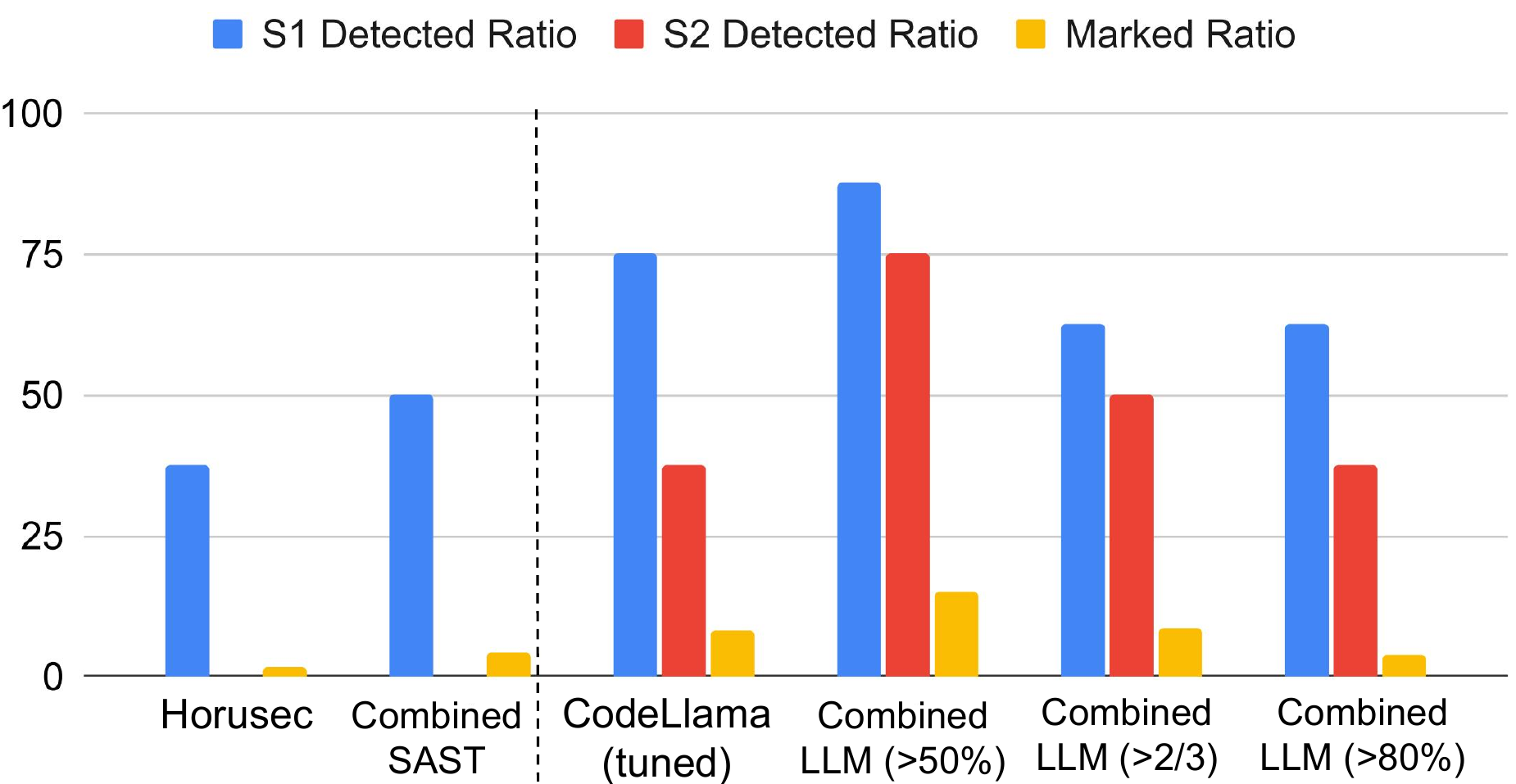}
    \vspace{-0.4cm}
    \caption{Combinations of SAST Tools or LLMs for Java}
  \label{fig:comb_java}
  \vspace{-0.3cm}
\end{figure}

\begin{figure}[t]
    \centering
    \includegraphics[width=0.42\textwidth]{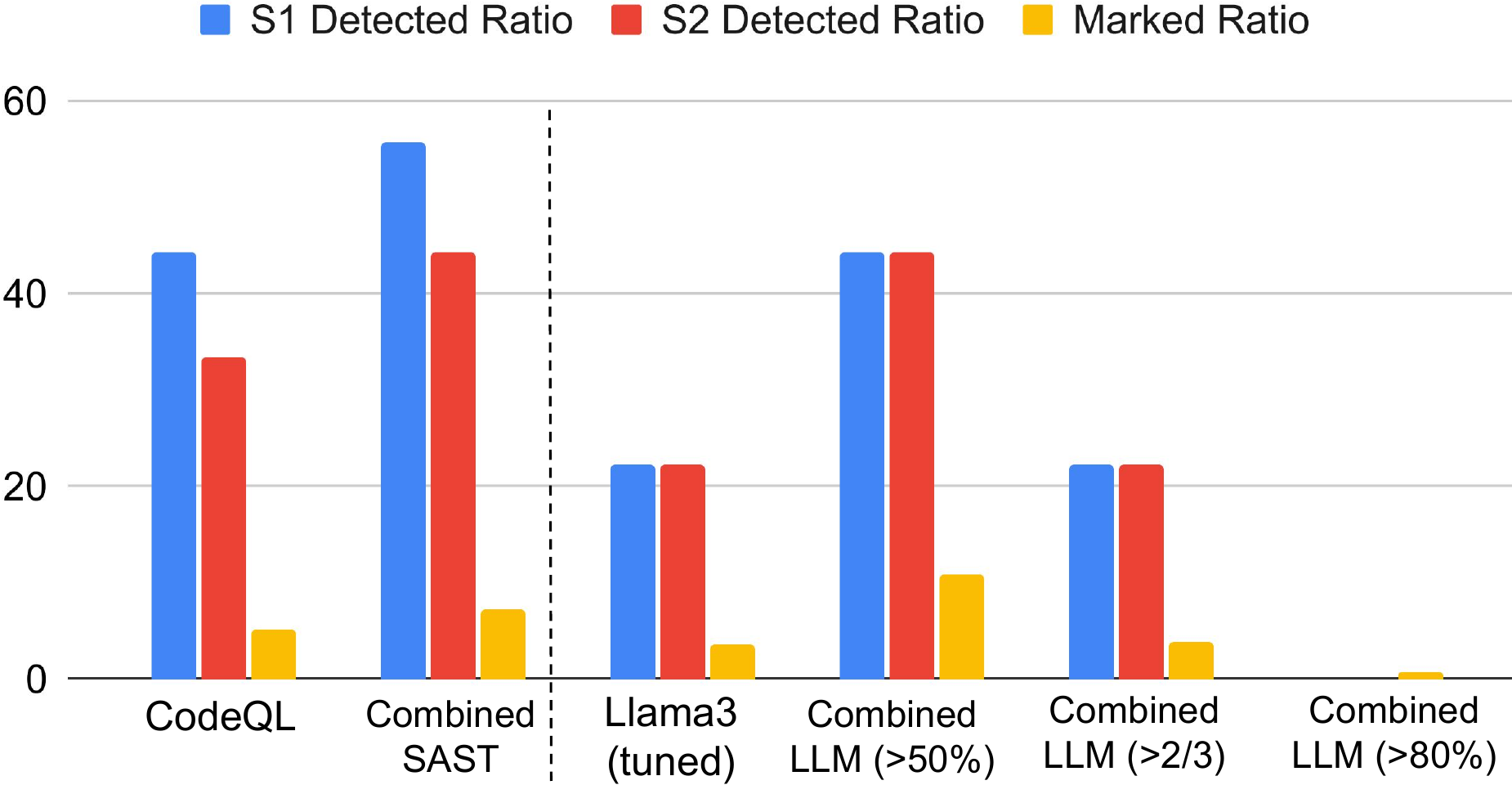}
    \vspace{-0.4cm}
   \caption{Combinations of SAST Tools or LLMs for C}
  \label{fig:comb_c}
  \vspace{-0.3cm}
\end{figure}

\begin{figure}[t]
    \centering
    \includegraphics[width=0.42\textwidth]{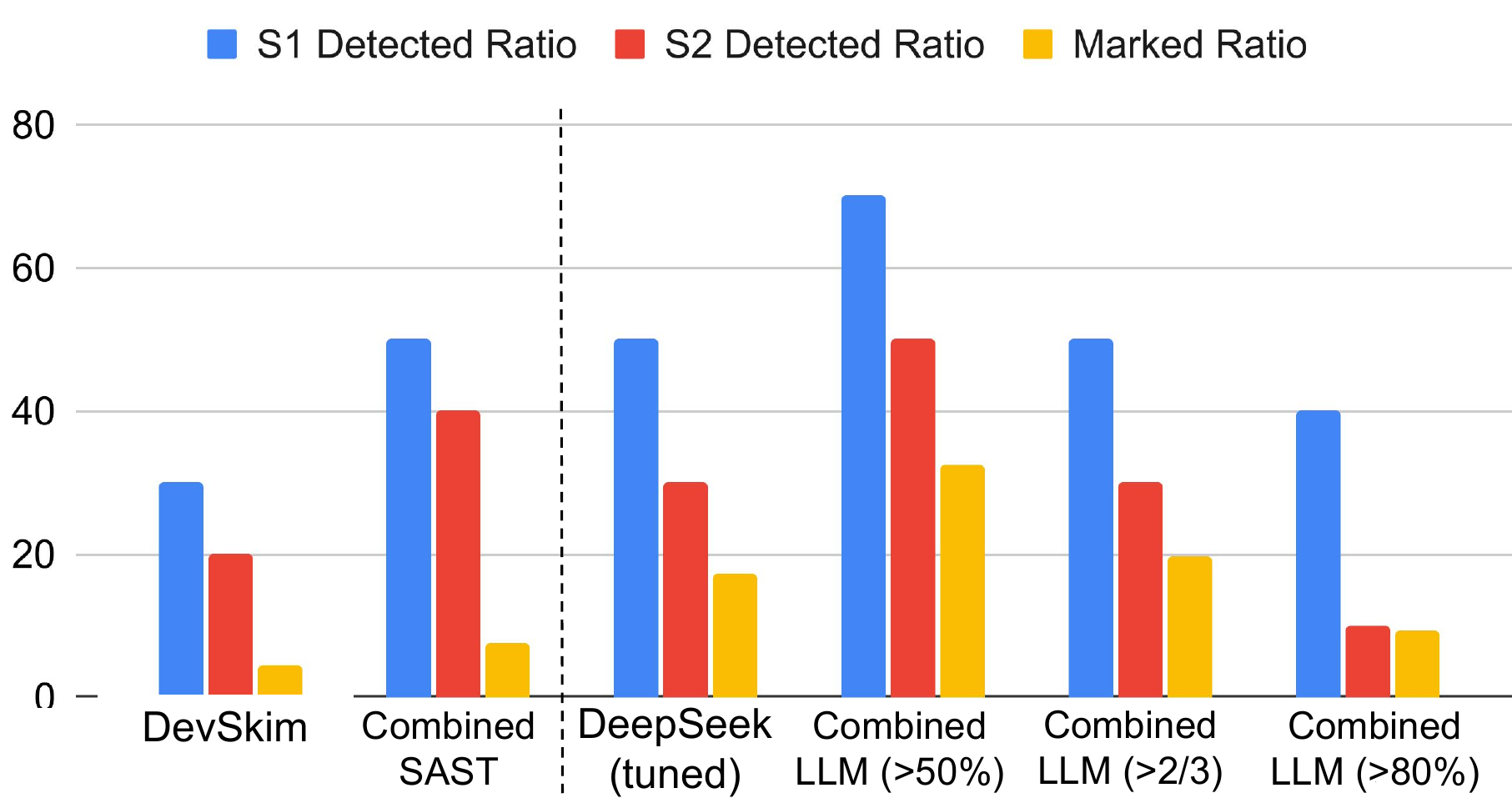}
    \vspace{-0.4cm}
    \caption{Combinations of SAST Tools or LLMs for Python}
  \label{fig:comb_python}
  \vspace{-0.5cm}
\end{figure}

The main limitation of SAST tools is their low detection ratios. Thus, the goal of combining SAST tools is to improve the detection ratios, even if it comes at the cost of higher marked function ratios.
To achieve this goal, we employ the following combining strategy to fuse the predictions of SAST tools: if a function is marked as vulnerable by any of the SAST tools, the function is considered to be marked as vulnerable by the SAST combination.
The first two columns of Figures ~\ref{fig:comb_java}--\ref{fig:comb_python} show the detection ratios and marked ratios of the individual SAST tool with the highest detection ratio, as well as the combination of all SAST tools, for Java, C, and Python, respectively.
\textbf{The results demonstrate that combining all SAST tools substantially boosts detection ratios, ranging from 25.2\% to 100.0\%.}
For Java, the detection ratio in Scenario 1 improved from 37.5\% to 50.0\%, at the cost of increasing the marked function ratio from 1.7\% to 4.3\%.
For C, the detection ratios improved from 44.4\% to 55.6\% in Scenario 1 and from 33.3\% to 44.4\% in Scenario 2, at the cost of increasing the marked function ratio from 5.2\% to 7.2\%.
For Python, the detection ratios improved from 30.0\% to 50.0\% in S1 and from 20.0\% to 40.0\% in S2, with an increase in the marked functions from 4.5\% to 7.7\%.

 \vspace{-0.1cm}
\subsubsection{Combining LLMs}
Different from SAST tools, the main limitation of LLMs is their high marked function ratios. Thus, the goal of combining LLM models is to reduce the marked function ratios, even if it comes at the cost of sacrificing some detection ratios.
Specifically, we include 12 LLMs in the LLM combination: four fine-tuned lightweight LLMs and eight fine-tuned large LLMs.
Regarding the combination strategies, we aim to reduce the number of marked functions. Thus, we employ three voting mechanisms on the predictions of 12 fine-tuned LLMs:
\textbf{1) C1 (voting>50\%)}: if more than half of the LLMs agree to mark one function as vulnerable, the combination of LLMs will mark the function as vulnerable;
\textbf{2) C2 (voting>$\frac{2}{3}$)}: if more than $\frac{2}{3}$ of LLMs agree to mark one function as vulnerable, the combination of LLMs will mark the function as vulnerable;
\textbf{3) C3 (voting>80\%)}: if more than 80\% of LLMs agree to mark one function as vulnerable, the combination of LLMs will mark the function as vulnerable.

The last four columns of Figures~\ref{fig:comb_java}--\ref{fig:comb_python} present the detection ratios and marked function ratios for one of the best individual LLMs (as determined from Figure~\ref{fig:tradeoff}), as well as for three different combinations of multiple LLM models for Java, C, and Python.
\textbf{The results show that combining multiple LLMs can substantially reduce the marked function ratios compared to individual LLMs on average.}
For instance, in the C2 LLM combination, it could reduce the marked function ratios by 61.7\%, 74.6\%, and 40.9\% in the Java, C, and Python benchmarks, respectively, compared to the average marked function ratios of the 12 fine-tuned LLMs. 
Additionally, the C2 LLM combination could achieve comparable detection ratios to the best-performing single LLM in each programming language (fine-tuned CodeLlama for Java, fine-tuned Llama3 for C, and fine-tuned DeepSeekCoder for Python), indicating that the detection ratios are not significantly compromised in this C2 combination.
\textbf{The more models we require LLMs to agree on the vulnerable predictions, the lower detection ratio and marked function ratios we can obtain.} For instance, when requiring agreement from more than 80\% of LLMs for a vulnerable prediction, the marked function ratios become comparable to those of SAST tools (0.8\%--9.2\%). However, this stringent agreement criterion leads to poor detection ratios, reaching only 0\%--37.5\% in Scenario 2.

\vspace{-0.1cm}
\subsubsection{Combined SAST v.s. Combined LLMs}
We discuss which combination strategy is better for each programming language.
\textbf{For Java, the C3 LLM combination is better} than the SAST combination because it has higher detection ratios and comparable marked function ratios.
\textbf{For C, the SAST combination is better} because it has higher detection ratios and a lower marked function ratio compared to the C1 LLM combination. In addition, the SAST combination obtains substantially higher detection ratios with an acceptable increase in marked function ratios compared to the C2 and C3 LLM combinations.
\textbf{For Python, the SAST combination is better.} Compared to the C2 and C3 LLM combinations, the SAST combination has comparable detection ratios with a much lower marked function ratio. Moreover, the marked function ratio of the C1 LLM combination is unacceptably high (i.e., 32.6\%), making it an unfavorable choice.

\vspace{0.2cm}
\noindent
\begin{tcolorbox} [boxrule=0.8pt,
                top=0.2pt,
                  bottom=0.2pt]
    \textbf{Answer to RQ3}: 
Combining SAST tools substantially boosts detection ratios, ranging from 25.2\% to 100.0\%.
Combining LLMs substantially reduces the marked function ratio by 40.9\%--74.6\% on average.
The optimal approach differs across programming languages.
For Java, the LLM combination provides the most effective solution, while for C and Python, the combination of SAST tools is better.
\end{tcolorbox}

\vspace{-0.2cm}
\section{Discussion}

 \subsection{Evaluating Commercial LLM ChatGPT}
In this work, we focus on open-source LLMs because using commercial LLMs for this task would be very costly, considering the large number of functions in each repository.
However, to make our paper more self-contained, we also conducted small-scale experiments with the famous commercial LLM ChatGPT. Specifically, we set the ratio of vulnerable functions to clean functions as 1:100 and sampled the corresponding number of clean functions from the repository. We used the \textit{gpt-3.5-turbo-0125} model~\cite{gpt35}, the latest version of GPT-3.5, and set the temperature to 0 for more reproducible results. Table~\ref{tab:gpt} presents the results of ChatGPT and two open-source LLMs, Llama3 and CodeBERT, with the same setting, averaged across Java, C, and Python. In general, the results show that the commercial LLM ChatGPT achieved lower vulnerability detection ratios and marked fewer functions as potentially vulnerable compared to the open-source LLMs such as Llama3 and CodeBERT. Among different ways of using ChatGPT, the few-shot prompting detects most vulnerabilities (i.e., 23.2\%).

\begin{table}[h]
\centering
\vspace{-0.2cm}
\caption{Evaluating ChatGPT on A Small Detection Range.}
\vspace{-0.4cm}
 \resizebox{0.49\textwidth}{!}{
\begin{tabular}{c|c|c|c|c|c}
\hline
\textbf{\begin{tabular}[c]{@{}c@{}}Pos:Neg\\ =1:100\end{tabular}} & \textbf{\begin{tabular}[c]{@{}c@{}}CodeBERT\\ (Tuned)\end{tabular}} & \textbf{\begin{tabular}[c]{@{}c@{}}Llama3\\ (Tuned)\end{tabular}} & \textbf{\begin{tabular}[c]{@{}c@{}}GPT3.5\\ (Zero-shot)\end{tabular}} & \textbf{\begin{tabular}[c]{@{}c@{}}GPT3.5\\ (CoT)\end{tabular}} & \textbf{\begin{tabular}[c]{@{}c@{}}GPT3.5\\ (Few-shot)\end{tabular}} \\ \hline
\textbf{S1 D $\uparrow$}                                                   & \textbf{89.2}                                                                & 67.4                                                              & 3.7                                                                   & 3.7                                                             &23.2                                                                      \\ \hline
\textbf{S2 D $\uparrow$}                                                   & \textbf{66.3}                                                                & 56.6                                                              & 3.7                                                                   & 3.7                                                             &11.1                                                                      \\ \hline
\textbf{Marked $\downarrow$}                                                 & 35.7                                                                & 20.1                                                              & \textbf{1.3}                                                                   & 2.1                                                             &10.8                                                                      \\ \hline
\end{tabular}
}
\label{tab:gpt}
\vspace{-0.3cm}
\end{table}

\vspace{-0.3cm}
\subsection{Implications}

\textbf{Repo-level vulnerability detection using LLMs holds substantial potential.}
While LLMs may not yet outperform SAST tools in every facet, as the first work exploring this direction, we have employed generic LLM techniques without incorporating specialized designs to maximize their effectiveness.
However, we have still observed that LLMs can achieve competitive performance compared to SAST tools, particularly when looking into individual SAST tools and LLMs.
This underscores the significant potential for advancing repo-level vulnerability detection through further development and refinement of LLM-based approaches.

\textbf{Repo-level vulnerability detection is more practical than the function-level vulnerability detection.}
We recommend a gradual shift in vulnerability detection efforts from the existing function-level task formation to the repo-level vulnerability detection task formation introduced in this study. As shown in Figure~\ref{fig:repo-fun-compare}, datasets used in traditional function-level vulnerability detection studies rely on vulnerability-fixing commits to narrow the detection scope from the entire repository to the specific files modified in those commits. However, when the goal is to detect vulnerabilities, the corresponding vulnerability-fixing commit does not yet exist. Consequently, it is impractical to depend on such commits to limit the detection range. 
The repo-level vulnerability detection formation addresses this issue by eliminating the need to narrow down the detection scope based on the fixing commits.

\textbf{Addressing the high marked function ratios of LLMs necessitates further investigation.} While LLMs tend to detect more vulnerabilities compared to SAST tools, they suffer from high marked function ratios, akin to high false positive rates. Although we managed to significantly reduce these ratios by combining multiple LLMs, there remains large room for enhancement. Potential solutions may entail developing techniques such as, filtering out less trustworthy predictions from LLMs, integrating additional contextual information (e.e., the structural features in the repository) with LLMs, refining prompting strategies, or employing more advanced ensemble methods.

\textbf{
Exploring advanced combinations of LLMs and SAST tools holds promise.} Each category of approach has its strengths and weaknesses in detecting vulnerabilities in software repositories: LLMs detect more vulnerabilities but with higher marked function ratios, whereas SAST tools detect fewer vulnerabilities but with lower marked function ratios. Future research could propose methods that can effectively leverage the strengths of both approaches for enhanced vulnerability detection.

\vspace{-0.3cm}
\subsection{Threats to Validity}

Threats to internal validity relate to possible errors and biases in our experiments. To mitigate these risks, we utilized the official implementations of the studied tools and models to ensure correctness. We have reviewed and validated our code and data, and they are publicly accessible to promote transparency and reproducibility. Another potential threat concerns the presence of undetected vulnerabilities in the vulnerability benchmarks. 
However, our objective was to assess whether the selected tools/models could accurately identify known vulnerabilities. Thus, it is reasonable to draw conclusions about the effectiveness of these tools/models in detecting known vulnerabilities.
Additionally, the potential for data leakage poses another threat, where LLMs might have been exposed to our evaluation datasets during their pre-training. This risk of data leakage is a widespread issue that affects numerous works related to LLMs (e.g., ~\cite{ahmed2024automatic,geng2024large}). Unfortunately, directly addressing this concern is almost impossible due to the non-public nature of pre-training datasets used by LLMs, especially commercial models like ChatGPT. 
Nevertheless, we observed that, compared to fine-tuning, LLMs do not perform effectively under zero-shot, chain-of-thought, and few-shot prompting. This suggests that memorization may not be a significant factor; if the models had already been pre-trained on our evaluation datasets, they would have achieved significantly better performance with the prompts, even without fine-tuning on our training sets.
Threats to external validity relate to the generalizability of our findings. To mitigate this potential concern, we employed a comprehensive list of 15 static analysis security testing tools and 12 popular or state-of-the-art large language models in this study. Additionally, we utilized experimental datasets spanning three popular programming languages: Java, C, and Python. Our extensive and diverse experiments help to alleviate concerns regarding the generalizability of our findings.

\vspace{-0.2cm}
\section{Related Work}

\noindent
\textbf{Studies of SAST Tools.}
There are many existing studies evaluating static analysis security testing (SAST) tools, e.g., ~\cite{C_CVE, Java_CVE,beba2019implementation,brito2023study,chen2020empirical, fan2018large, goseva2015capability, habib2018many,kang2022detecting,liu2023comprehensive,nguyen2021adoption,tomassi2018bugs,kaur2020comparative,kaur2020comparative,goseva2015capability,thung2012extent,nunes2019empirical,qiu2018analyzing,alqaradaghi2021detecting,li2019evaluation}. Some of these studies utilize synthetic benchmarks to evaluate SAST tools, e .g., ~\cite{alqaradaghi2021detecting,beba2019implementation,li2019evaluation}. Several studies have evaluated SAST tools on real-world vulnerability datasets in C/C++ or Java~\cite{C_CVE,Java_CVE,kaur2020comparative,goseva2015capability,thung2012extent}. For instance, Lipp et al.~\cite{C_CVE} focused on SAST tools for C programs, evaluating the effectiveness of six tools on real-world vulnerability datasets. 
Li et al.~\cite{Java_CVE} evaluated the effectiveness of seven free or open-source SAST tools on real-world vulnerability datasets.
Several studies have evaluated SAST tools in other research areas, such as Android~\cite{chen2020empirical,qiu2018analyzing,fan2018large}.

Our work distinguishes itself from existing studies in the following aspects. Firstly, we evaluated SAST tools and LLM models on three popular programming languages: Java, C, and Python. This ensures diversity in the experimental datasets, unlike many previous studies that focused on a single programming language.
Secondly, to the best of our knowledge, we are the first to evaluate SAST tools for Python vulnerability detection. 
Thirdly, we are the first to comprehensively compare SAST tools with LLMs, which have demonstrated strong effectiveness in many other code-related tasks such as code generation~\cite{liu2024your} and automated repair~\cite{jiang2023impact}.

\vspace{0.1cm}
\noindent
\textbf{LLM-based Vulnerability Detection.}
There are many existing studies utilizing large language models (LLMs) for vulnerability detection, e.g., ~\cite{codebert,graphcodebert,DBLP:conf/icse/YangWLW23,liu2024pre,DBLP:conf/scam/PengCZTLH23,DBLP:journals/tse/ZhangLHXL23,DBLP:conf/ijcnn/HanifM22,DBLP:journals/corr/abs-2401-15468,fu2023chatgpt}.
For instance, Liu et al.~\cite{liu2024pre} leveraged the abstract syntax tree and program dependence graph of functions to pre-train their LLM. The pre-training process enabled the LLM to predict dependencies within the functions effectively.
Peng et al.~\cite{DBLP:conf/scam/PengCZTLH23} employed program slicing, a technique that extracts control and data dependency information, to enhance the capability of LLMs in vulnerability detection.
Additionally, Zhang et al.~\cite{DBLP:journals/tse/ZhangLHXL23} proposed an approach that involved decomposing syntax-based control flow graphs into multiple execution paths. These execution paths were then fed as input to LLMs for enhancement.
Recently, a parallel work~\cite{wen2024vuleval} proposed an approach to enhance the function-level vulnerability detection task by retrieving relevant dependencies from the repository.  
Specifically, their approach involves concatenating the target functions (those in files changed by vulnerability-fixing commits) with the relevant dependencies retrieved from the repository. 
In contrast to their work, our work aims to assess and compare the effectiveness of SAST tools and LLMs in detecting vulnerabilities from entire repositories rather than from files changed by the fixing commits.

While numerous studies have explored the use of LLMs for vulnerability detection, they have focused on identifying vulnerabilities within individual functions. In addition, existing works have not comprehensively compared their approaches with diverse SAST tools, and they have typically focused on one programming language. Our research diverges from previous works by (1) evaluating the efficacy of LLMs in detecting vulnerabilities from entire repositories (repo-level vulnerability detection), and (2) evaluating their efficacy against a comprehensive list of 15 SAST tools, spanning three widely-used programming languages: Java, C, and Python.

\vspace{-0.2cm}
\section{Conclusion and Future Work}
In this study, we compare the efficacy of 15 diverse SAST tools alongside 12 popular or leading LLMs. 
Our experiments focus on real-world vulnerabilities across three prevalent programming languages: Java, C, and Python.
Our findings indicate that SAST tools exhibit low vulnerability detection ratios while maintaining a low marked function ratio, akin to a low false positive rate. 
In contrast, LLMs demonstrate high vulnerability detection ratios but are accompanied by elevated marked function ratios (akin to high false positive rates).
Through ensemble approaches, we demonstrate that combining SAST tools and LLMs mitigates their respective limitations, resulting in improved overall vulnerability detection performance.

In the future, we aim to investigate more sophisticated methods for effectively combining multiple LLMs and SAST tools to further improve vulnerability detection performance.

\balance
\bibliographystyle{ACM-Reference-Format}
\bibliography{Citation}


\end{document}